\def\hh{\hbox{$^{\rm h}$}}
\def\mm{\hbox{$^{\rm m}$}}
\def\am{\hbox{$^{\prime}$}}
\documentstyle[12pt,aaspp4,psfig,flushrt]{article}
\begin{document}

\title{THE SYMBIOTIC NEUTRON STAR BINARY GX~1+4/V2116~OPHIUCHI}  

\author{Deepto~Chakrabarty\altaffilmark{1}}
\affil{\footnotesize Center for Space Research, Massachusetts
Institute of Technology, Cambridge, MA 02139; deepto@space.mit.edu}

\and

\author{Paul~Roche}
\affil{\footnotesize Astronomy Centre, University of Sussex, Brighton
BN1 9QH, England, UK; pdr@star.maps.susx.ac.uk}

\altaffiltext{1}{NASA Compton GRO Postdoctoral Fellow}

\medskip
\centerline{Submitted March 10, 1997; revised May 30, 1997}
\centerline{Accepted for publication in {\sc The Astrophysical Journal}}

\begin{abstract}

We present extensive optical, infrared, and X-ray observations of the
S-type symbiotic low-mass X-ray binary GX 1+4/V2116 Oph, which
consists of a 2-min X-ray pulsar accreting from an M6 III giant. This
is the only symbiotic system definitely known to contain a neutron
star. The mean observed spectral type of the X-ray-heated mass donor
is M5 III. The steady interstellar extinction toward the binary
($A_V=5.0\pm0.6$) contrasts the variable hydrogen column density
inferred from X-ray measurements, most likely evidence for a variable
stellar wind. The mass donor is probably near the tip of the
first-ascent red giant branch, in which case the system is 3--6 kpc
distant and has an X-ray luminosity of $\sim 10^{37}$ erg s$^{-1}$. It
is also possible, though less likely, that the donor star is just
beginning its ascent of the asymptotic giant branch, in which case the
system is 12--15 kpc distant and has an X-ray luminosity of $\sim
10^{38}$ erg s$^{-1}$. However, our measured $A_V$ argues against such
a large distance.

We show that the dense ($N_e\sim 10^9$ cm$^{-3}$) emission-line nebula
enshrouding the binary is powered by ultraviolet radiation from an
accretion disk. The emission-line spectrum constrains the temperature
profile and inner radius of the disk (and thus the pulsar's magnetic
field strength), and we mention the implications this has for
explaining the accretion torque reversals observed in the 
pulsar.  We also show that the binary period must be $\gtrsim 100$ d
and is most likely $\gtrsim 260$ d, making GX 1+4 the only known
low-mass X-ray binary with $P_{\rm orb}> 10$ d.  If the mass donor
fills its Roche lobe, the mass transfer rate must be highly
super-Eddington, requiring considerable mass loss from the binary.  We
discuss the alternative possibility that the accretion disk forms from
the slow, dense stellar wind expected from the red giant.

\end{abstract}


\section{INTRODUCTION}

Low-mass X-ray binaries (LMXBs) with evolved giant donors are the
current state or probable progenitor of most low-mass neutron star
binaries with wide ($P_{\rm orb} \gtrsim 2$ d) orbital separations
(see Verbunt \& van den Heuvel 1995 for a review).  In the usual scenario,
mass transfer onto the neutron star begins when its low mass companion
ascends the giant branch and fills its Roche lobe. This mass transfer,
driven by the nuclear evolution and expansion of the giant, continues
at a high rate until the donor envelope is exhausted, eventually
leaving a wide white dwarf/neutron star binary.  Tidal torques exerted
by the red giant envelope act rapidly to circularize the orbit.

Many examples of systems which probably evolved in this way are
known. Among current LMXBs, Cyg X-2 and 2S 0921--63 are in long-period
($\sim$ 9 d) orbits with moderately evolved (F or G type) giants, and
the recently discovered bursting X-ray pulsar GRO J1744--28 is in a
wide, 11.8-d circular orbit suggestive of tidal interactions with a
giant companion (Finger et al. 1996).  There are over 30 known
low-mass binary radio pulsars in wide circular orbits whose white
dwarf companions are probably the remnant core of a red giant whose
envelope was exhausted or ejected during mass transfer (Phinney \&
Kulkarni 1994; Camilo 1995).  However, some of these radio pulsar
binaries have orbital periods much longer ($P_{\rm orb}\gg 10$~d) than
the widest known LMXBs.  The presumed progenitors of such systems,
LMXBs accreting from highly evolved (K or M type) giant donors, have
not been previously identified.  They are expected to be rare, owing
to the short duration of both the K/M giant lifetime and the highly
unstable mass transfer stage expected for Roche-lobe-overflow LMXBs
with $P_{\rm orb}\gtrsim 2$~d (Kalogera \& Webbink 1996).  The high
mass transfer rate might even smother the X-ray emission, making the
system unobservable as an LMXB.

If such a system were observed, it would almost certainly be
classified as symbiotic binary.  These consist of a K/M giant primary
and a hot secondary enshrouded by an emission line nebula, which
results from photoionization of the red giant's wind by the secondary
(see Kenyon 1986 and Iben \& Tutukov 1996 for reviews). Their
observational characteristic is a composite spectrum in which strong
emission lines are superimposed on the cool molecular absorption
spectrum of the highly evolved primary. Over 100 symbiotic binaries
are known, most of which have hot main sequence or white dwarf
secondaries.  The only symbiotic known to have a neutron star
companion is the X-ray pulsar binary GX 1+4/V2116 Oph (Davidsen,
Malina, \& Bowyer 1977), although the X-ray binary 4U 1700+24/HD
154791 (Garcia et al. 1983) may be a second example.  No other LMXBs
with M giant companions are known.  GX 1+4 is thus a promising
candidate progenitor for the widest low-mass neutron star binaries
(e.g., Taam \& van den Heuvel 1986).

The $\sim 2$~min accretion-powered binary pulsar GX 1+4 (=
4U~1728--247) was discovered in a hard X-ray balloon observation over
25 years ago (Lewin, Ricker, \& McClintock 1971). Throughout the
1970s, it was the brightest hard X-ray source in the Galactic center
region ($\sim$50--200~mCrab), and the pulsar was spinning up at a
rapid rate ($|\nu/\dot\nu|\approx 40$~yr).  In the early 1980s GX 1+4
entered an extended low state, with soft X-ray flux of $<0.5$~mCrab
(Hall \& Davelaar 1983; Mukai 1988). When it reemerged with low
luminosity ($\sim$~2~mCrab) in 1987, the pulsar was spinning down
rapidly at nearly the same rate that it had been spinning up
previously (Makishima et al. 1988). GX 1+4 spun down and gradually
brightened to $\sim$100~mCrab until 1994, when it briefly resumed
rapid spin-up before returning to a spin-down state (Chakrabarty et
al. 1997).  If interpreted in terms of standard magnetic
accretion torque theory for X-ray pulsars accreting from disks, (e.g.,
Ghosh \& Lamb 1979), these torque reversals imply an extraordinarily
strong ($\sim 10^{14}$~G) dipole magnetic field at the surface of the
neutron star. 

Shortly after its discovery, GX 1+4 was identified with the bright
infrared source V2116 Oph (Glass \& Feast 1973).  Optical spectroscopy
of this star clearly established it as a symbiotic (Davidsen et
al. 1977).  In addition to a number of very bright emission lines due
to H I (especially H$\alpha$) and He I, the optical spectrum also
contained emission lines of several highly ionized species, notably
[\ion{Fe}{7}] and [\ion{Fe}{10}]. This indicated the presence of
high-energy ionizing photons, suggesting that the hot secondary was a
compact object and thus strengthening the association with GX 1+4.
Optical pulsations consistent with the spin period of GX 1+4 have
recently been detected from V2116 Oph (Jablonski et al. 1997).

In this paper, we present extensive observations of the GX 1+4/V2116
Oph system at optical, infrared, and X-ray wavelengths. 

\section{OBSERVATIONS}

\subsection{X-ray observations}

The pulsed hard X-ray (20--60 keV) emission from GX 1+4 has been
observed continuously since 1991 April with the BATSE all-sky monitor
on the {\em Compton Gamma Ray Observatory}. These observations are
described in detail elsewhere (Chakrabarty et al. 1994, 1997).
However, to provide context for the multiwavelength observations
presented in this paper, Table 1 lists the 20--60 keV pulsed intensity
of GX 1+4 measured by BATSE for the times of each of the post-1991
multiwavelength observations described in this paper.  The hard X-ray
pulsed intensities are excerpted from a BATSE pulsed flux history
given by Chakrabarty (1996). It should be emphasized that these
intensities represent only the pulsed component of the emission above
20 keV. Although the extent to which the BATSE pulsed intensity is a
precise tracer of the bolometric flux from GX 1+4 remains an open
question (Chakrabarty et al. 1997), we will treat it as an approximate
tracer in this paper.

We obtained two 0.1--2.4 keV observations of the GX 1+4 field with the
{\em ROSAT} High Resolution Imager (HRI). The first observation was
for a total of 4894 s on 1995 March 12--16.  No point sources were
detected in the field of view. The 95\%-confidence upper limit on
emission from GX 1+4 was 0.0029 count s$^{-1}$. 

A second observation was made for a total of 5861 s on 1996 March
26--28. A total of $305\pm17$ counts was detected from GX 1+4,
corresponding to an intensity of $0.0506\pm 0.0030$ count s$^{-1}$
with a pulsed fraction of approximately 70\%. The derived coordinates
of GX 1+4 with respect to the J2000.0 equinox are: RA =
17\hh\,32m\,02\fs31, Dec = $-24^\circ$\,44\am\,44\farcs10. The
90\%-confidence error circle due to photon statistics and the
point-spread function of the instrument has radius
0.8 arcsec. There is an additional 6.3 arcsec (1$\sigma$) systematic
uncertainty in the {\em ROSAT} attitude solution, resulting in an
overall 90\%-confidence error circle radius of 10.4 arcsec. The center
of our error circle lies within 4 arcsec of the optical coordinates of
V2116 Oph.  It also lies within 3 arcsec of the center of an earlier
11.4-arcsec-radius (90\%-confidence) {\em ROSAT}/PSPC position
determined from only 15 counts (Predehl, Friedrich, \& Staubert
1995). The azimuthally-averaged surface brightness profile of the GX
1+4 region shows no evidence for the faint, diffuse halo reported by
Predehl et al. (1995).

\subsection{Optical astrometry and optical/IR photometry}

Optical astrometry of the GX 1+4 field was obtained using plate
J2333 from the SERC(J) southern sky survey, exposed at the UK Schimdt
Telescope in Australia on 1976 May 27. Astrometric coordinates for
V2116 Oph and several neighboring stars, precessed to the J2000.0
equinox, are given in Table 2. An $R$-band finder image of the field,
shown in Figure 1, was acquired on 1993 September 10 with the $f/11$
Nasymth EMMI camera on the 3.5-m New Technology Telescope (NTT) at the
European Southern Observatory (ESO) in La Silla, Chile. The 90\%
confidence {\em ROSAT} error circle derived above for the X-ray source
is also shown.

Deep optical CCD photometry was obtained on 1993 June 29 using the
$f/2.8$ prime focus COSMIC camera on the 5-m Hale telescope at Palomar
Mountain, California. The $B$ (4400 \AA), $V$ (5500 \AA), and $R$
(7000 \AA) magnitudes of V2116 Oph are given in Table 2. To aid future
differential photometry, the magnitudes for several field stars are
given as well. Infrared photometry in the $J$ (1.25 $\mu$m), $H$ (1.65
$\mu$m), and $K$ (2.2 $\mu$m) bands was obtained during 1993--1995
using the the $f/50$ Cassegrain Mk III infrared photometer on the SAAO
1.9-m Radcliffe telescope; the infrared camera (IRCAM) on the 3.8-m
United Kingdom Infrared Telescope (UKIRT) at Mauna Kea, Hawaii; and
the continuously-variable filter photometer (CVF) on the 1.5-m
Telescopio Carlos Sanchez (TCS) at the Teide Observatory in Iza\~na,
Tenerife, Canary Islands, Spain. In some of the SAAO observations, $L$
(3.6~$\mu$m) band measurements were also acquired, and several $L'$
(3.7~$\mu$m) band measurements were obtained during the UKIRT
observations.  The complete infrared photometric history of V2116 Oph,
including all previously published measurements, is summarized in
Table 3.

\subsection{Optical/IR spectroscopy}

Spectroscopic observations of V2116 Oph were obtained on 1991 August
8--9 and 1993 June 30 using the $f/15.7$ Cassegrain double
spectrograph of the Palomar 5m Hale telescope. 
A service observation was made during a bright X-ray outburst of GX 
1+4 on 1993 September 10, using the $f/11$ Nasmyth EMMI spectrograph
on the 3.5-m New Technology Telescope (NTT) of the European Southern
Observatory (ESO) in Cerro La Silla, Chile. This observation, made
with EMMI in its red imaging and low dispersion (RILD) mode, was
acquired during an engineering night, and a significant part of the beam
was obscured leading to a rather low count rate; however, the
obscuration was identical for all the data acquired.  

Several archival observations are available from the 3.9-m
Anglo-Australian Telescope (AAT) at Siding Spring Mountain, New
South Wales, Australia. Spectra were acquired on 1976 April 10
and 1976 August 20  using the Boller and Chivens spectrograph and the
IDS detector.  The 1976 August spectrum was previously published by
Whelan et al. (1977). An additional observation was obtained on 1984
April 22, during the X-ray low state of GX 1+4 reported by {\em
EXOSAT}.  More recently, two AAT service observations were made
on 1993 September 10 using the $f/8$ Cassegrain RGO spectrograph. 
We obtained a high-resolution H$\alpha$ observation with this
spectrograph on 1994 February 26. 

An H$\alpha$ spectrum was also obtained on 1993 July 1
using the $f/11$ ISIS double spectrograph on the 4.2-m William
Herschel telescope at the Roque de los Muchachos Observatory in La
Palma, Canary Islands, Spain. Several additional H$\alpha$
observations were obtained during 1993--1995 using the $f/18$
Cassegrain image tube spectrograph and Reticon detector on the 1.9-m
Radcliffe telescope of the South African Astronomical Observatory
(SAAO) in Sutherland, Cape Province, South Africa. 

All of the optical spectra were reduced using the Figaro analysis
package (Shortridge 1991). Where suitable spectrophotometric standards
were observed, the data were flux-calibrated and corrected for
atmospheric extinction.  We corrected the observed wavelengths for the
Earth's motion with respect to the solar system barycenter.

A summary of all known optical spectroscopic observations of V2116 Oph,
including previously published measurements, is given in Table 4.
Representative wide coverage spectra are shown in Figure 2.  A more
detailed medium-resolution spectra is shown in Figure 3. 

Low-resolution ($\lambda/\Delta\lambda\sim150$) infrared grism
spectroscopy of V2116 Oph in the $JHK$ bands were obtained on 1993
April 6 with the $f/70$ Cassegrain infrared camera and a 58$\times$62
InSb array on the Palomar 5-m Hale telescope. These spectra were
obtained using a 1 arcsec slit and broad-band {\em JHK} filters for order
sorting. Instrumental and atmospheric absorption features were removed
using the featureless continuum of the G0\,V star HR 4345 (see Graham
et al. 1992 for details).

\section{RESULTS}

\subsection{Spectral classification}

Davidsen et al. (1977) assigned a tentative classification of M6\,III
to V2116 Oph, warning that ``in view of the complicated nature of the
spectrum, this preliminary estimate should be accepted with caution.''
We have several available spectra which include the
$\lambda\lambda$7300--8100 range and thus are suitable for checking
the classification. A visual comparison of the near-infrared spectra
in Figure 2 with the M giant spectral sequences of Kirkpatrick et
al. (1991) confirm that the star is mid to late M class, in particular
by means of the TiO and VO absorption bands, which are sensitive to
temperature in M giants.  Kenyon \& Fernandez-Castro (1987) and
Terndrup et al.  (1990) have defined absorption indices to measure the
depth of these features relative to the interpolated continuum. The
$I(6180)$=[TiO]$_1$ and $I(7100)$=[TiO]$_2$ indices of Kenyon \&
Fernandez-Castro (1987) and the $S(7890)$ index of Terndrup et al.
(1990) measure the depth of the TiO features, while the $I(7865)$=[VO]
(Kenyon \& Fernandez-Castro 1987) and $I(7450)$ (Terndrup et al. 1990)
indices measure VO absorption. All of these indices have been
extensively calibrated against spectral type using a large number of K
and M giants and supergiants and are relatively insensitive to reddening.

When measuring these indices in V2116 Oph, we found that the two TiO
indices of Kenyon \& Fernandez-Castro gave spurious results, measuring
negligible TiO $\lambda\lambda$6180, 7100 absorption (typical of K and
early M stars) despite the obvious presence of stronger features at
longer wavelengths. Since this is probably due to contamination of the
red giant absorption spectrum by nebular emission or an accretion
disk, we confined our analysis to the other three absorption indices
at longer wavelengths, which all give consistent results (see Table
5). The 1975 August, 1984 April, and 1993 June spectra
are all consistent with an M5 star, but the 1993 September 10 spectrum
indicates a somewhat higher temperature, around M3.

CO absorption strength in late type stars is correlated with both
luminosity and effective temperature (e.g., Kleimann \& Hall
1986). Doyon et al. (1994) quantified this relationship in terms of a
spectroscopic index for CO absorption at 2.3 $\mu$m. Using our
measurement of this feature in the 1993 April Palomar data (Figure 4),
we find [CO]$_{\rm sp}$=0.31, which clearly identified V2116 Oph as
luminosity class III.  High-resolution infrared spectroscopy of the CO
absorption band features may be able to place more quantitative
constraints on the luminosity.

X-ray heating is probably responsible for our M3 measurement on 1993
September 10. Data from the BATSE all-sky monitor on the {\em Compton
Gamma Ray Observatory} indicate that the hard X-ray (20--60 keV)
pulsed flux from GX 1+4 on 1993 September 10 was a factor of 3 higher
than on 1993 June 30 (see Table 1). If we assume that the BATSE
measurements are a tracer of the bolometric luminosity of GX 1+4 and
that the temperature change is due to a change in the X-ray heating
of the red giant photosphere between the two observations, then we can
write
\begin{equation}
T_{\rm M3}^4  - \frac{3 L_x f \alpha}{2\pi\sigma R_g^2}
  = T_{\rm M5}^4 - \frac{ L_x f \alpha}{2\pi\sigma R_g^2},
\end{equation}
where $T_{\rm M3}$=3675 K and $T_{\rm M5}$=3470 K are the effective
temperatures for the red giant (Dyck et al. 1996), $R_g$ is the
red giant radius, $L_x$ is the X-ray luminosity in
``quiescence'' (that is, on 1993 June 30), $\alpha$ is the X-ray
albedo of the red giant, $f$ is the fraction of $L_x$
intercepted by the red giant surface, and $\sigma=5.67\times 10^{-5}$
erg cm$^{-2}$ K$^{-4}$ s$^{-1}$ is the Stefan-Boltzman constant.  For
a neutron star/red giant binary, we estimate $f\approx 0.03$ (e.g.,
Webbink, Rappaport, \& Savonije 1983). Then, assuming $\alpha=0.5$, we
find that $L_x = 10^{37}\,R^2_{100}$ erg s$^{-1}$, where
$R_{100}$ is the red giant radius in units of $100 R_\odot$.  This
calculation suggests that the effective temperature of the {\em
unheated} red giant is $\approx 3370$ K (corresponding to an M6 III
star; Dyck et al. 1996), even though the mean {\em observed} spectral
type is M5 III.   However, the calculation should be treated with
caution; the X-rays may only ionize the red giant's atmosphere without
penetrating to the photosphere (see, e.g., Proga et al. 1996).
Further correlative study of the M giant's temperature and optical
line profile variations with changing X-ray illumination may help
clarify this point. 

As a crude consistency check on our classification, we can use our
infrared photometry to compute a color temperature. Fitting our 1993
April {\em JHKL} photometry to a reddened ($A_V=5$; see below)
blackbody, we find $T_{\rm color} = 2590$ K, which corresponds to an
M6 III star (Ridgway et al. 1980). The resulting blackbody curve is
compared with our low-resolution infrared spectra in Figure 4.  The
significant difference between our estimates of $T_{\rm color}$ and
$T_{\rm eff}$, as well as the strong $H$-band excess above a reddened
blackbody curve, can be understood in terms of the complicated
near-infrared molecular opacities in the atmospheres of cool stars
(Tsuji 1966; Wing 1981).

Infrared photometry may also be used to further classify symbiotic
stars into one of two groups: the S-type (``stellar'') symbiotics,
whose IR colors are consistent with a late type giant; and the D-type
(``dusty'') symbiotics, where the cool star is usually a Mira variable
and whose IR colors indicate the presence of even cooler dust emission
(Webster \& Allen 1975; Allen 1982; Munari et al. 1992).  The {\em
JHKL} colors of V2116 Oph fall close to the boundary of the two
groups, but are most consistent with an S-type classification. 

\subsection{Reddening and column density}

The optical colors of V2116 Oph are clearly inconsistent with those of
an M giant for any amount of interstellar reddening, suggesting that
much of the optical continuum is dominated by disk or nebular
emission. Moreover, the \ion{H}{1} Balmer and \ion{He}{1} line ratios
are also inconsistent with recombination in a standard ``case B''
radiative nebula (i.e. optically thick in the \ion{H}{1} Lyman lines;
Baker \& Menzel 1938; Osterbrock 1989) for any amount of reddening, as
is often seen in regions of high density and optical depth
(Osterbrock 1989). Following Davidsen et al. (1977), we instead employed
the infrared colors to determine the reddening.

Care must be taken to measure both the spectral type and the infrared
colors simultaneously, since the temperature-dependence of M giant
intrinsic colors on a $JHK$ color-color diagram (Bessell \& Brett
1988) parallels the interstellar reddening vector (Rieke \&
Lebofsky 1985).  Fortunately, on two of the dates for which we have
spectral classifications (1993 June 30 and 1993 September 10), we also
obtained $JHK$ photometry.  Comparing the measured $JHK$ colors on
these dates with the intrinsic colors for M5 III and M3 III giants,
respectively (Bessell \& Brett 1988), and assuming the reddening is
constant over time, we find mean infrared color excesses of $\langle
E(J-H)\rangle=0.53\pm 0.03$ and $\langle E(H-K)\rangle=0.33\pm
0.03$. Then, using the interstellar reddening law of Rieke \& Lebofsky
(1985), we infer that $E(B-V)=1.63\pm 0.19$, $A_V= 5.0\pm 0.6$, and
$A_K=0.56\pm 0.08$.  The error bars for the last three quantities
include the systematic uncertainty in the interstellar extinction law
in addition to the measurement errors.

The stability of the infrared colors indicates that the observed
extinction is constant, and thus probably interstellar in origin.
Using the empirically measured relationship between optical extinction
and interstellar X-ray absorption (Gorenstein 1975), we would expect a
similarly constant hydrogen column density to the source of $N_{\rm
H}\approx 1\times 10^{22}$ cm$^{-2}$.  However, archival X-ray
spectral measurements of the column density have found substantial
variability (e.g. Becker et al. 1976), but with a typical value of
$N_{\rm H}\approx 4\times 10^{22}$ cm$^{-2}$. At the same time, the
inferred unabsorbed X-ray continuum shape has been fairly constant.
It is interesting to note that our two {\em ROSAT} observations of
GX~1+4, separated by a year, found an order of magnitude difference in
the 0.1--2.4 keV intensity despite a difference of only 25\% in the
20--60 keV intensity measured simultaneously by BATSE. The flux
measured in the {\em ROSAT} band is extremely sensitive to the
hydrogen column density. Our 1996 March {\em ROSAT} measured intensity
is consistent with the level using the PIMMS simulation
program\footnote{PIMMS (Portable, Interactive, Multi-Mission
Simulator) is a tool for predicting count rates in X-ray astronomy
missions for a variety of source spectral models. It was written by
Koji Mukai at NASA/Goddard Space Flight Center and is available on the
World Wide Web via anonymous FTP at
{\tt ftp://legacy.gsfc.nasa.gov/software/tools}.}
assuming the column density ($N_{\rm H}=4\times 10^{22}$ cm$^{-2}$)
and spectral shape measured by {\em Rossi X-Ray Timing Explorer} in
1996 February (W. Cui 1996, priv. communication), and using the
simultaneous BATSE 20--60 keV measurement for normalization. The 1995
March {\em ROSAT} non-detection is consistent with a higher column
density $\gtrsim 7\times 10^{22}$ cm$^{-2}$, well within the range of
observed column density variability in GX 1+4. We conclude that the
observed optical extinction is mainly due to interstellar dust, while
the variable X-ray absorption is caused by gas local to GX 1+4,
probably the wind of the red giant.

\subsection{Emission line spectrum}

We used our extinction estimate to compute the dereddened emission
line fluxes $I(\lambda)=F(\lambda) e^{A(\lambda)/1.086}$ for each of
our observations, where $F(\lambda)$ is the observed emission line
flux and $A(\lambda)$ is the extinction in magnitudes as a function of
wavelength. To compute $A(\lambda)$, we fit a cubic polynomial to the
interstellar extinction law of Rieke \& Lebofsky (1985):
$[A(\lambda)/A_V] = 0.01472 + 0.01425 (1/\lambda) + 0.49011
(1/\lambda)^2 - 0.10709 (1/\lambda)^3$, where the wavelength $\lambda$
is measured in microns. This fit is valid over the range 0.3 $\mu$m
$<\lambda<$ 5 $\mu$m. The resulting dereddened emission line fluxes
are included in Tables 6, 7, and 8.  Table 6 contains a selected line
list for the archival spectra from 1974--1988 as well as the 1993 June
Palomar spectrum. Tables 7 and 8 contains detailed line lists for our
observations on 1991 August 8 and 1993 September 10, respectively.
The optical emission line features were identified using the tables of
Meinel, Aveni, \& Stockton (1975).

The dominant feature in all of the optical spectra is a tremendous
H$\alpha$ emission line. \ion{He}{1} lines are also detected in all of
the spectra. Very few forbidden lines are observed, indicating a high
electron density in the emission line region. The 1991 August data
from Palomar (Figure 3) provide the first high signal-to-noise
measurement of the blue-end spectrum of V2116 Oph, revealing several
hydrogen Balmer lines as well as weak emission from
\ion{He}{2} $\lambda$4686 and numerous other species.  Three spectra
were obtained on 1993 September 10 during a bright X-ray flare
from GX 1+4. The ESO/NTT spectrum (bottom panel of Figure 2) extends
into the near infrared, measuring the strong \ion{Ca}{2} triplet and
the extraordinarily strong \ion{O}{1} $\lambda$8446 line, as well as a
number of weaker lines from the hydrogen Paschen series.  In the
low-resolution infrared spectra (Figure 4), the H I Pa$\beta$ and
Br$\gamma$ lines are detected.

A broad emission feature near $\lambda$6830 is present in the
older spectra (1974--1988) and is weakly detected in the 1991 August
spectrum. This feature, which is believed to arise from Raman
scattering of \ion{O}{6} $\lambda\lambda$1032, 1038 by neutral
hydrogen (Schmid 1989), is in general only observed in those symbiotic
stars which have highly excited species like [\ion{Fe}{7}] 
(Allen 1980). Indeed, the [\ion{Fe}{7}] features reported by Davidsen
et al. (1977) and detected in the 1976--1988 spectra are undetected in
the more recent (1991--1995) spectra, consistent with the
correspondingly weaker $\lambda$6830 emission. 

The presence of numerous \ion{Fe}{2} lines combined with the absence
of any \ion{Fe}{3} features indicates that the electron temperature
$T_e<30000$ K (Osterbrock 1989), so that the rich emission line
spectrum we observe must be due to photoionization rather than
collisional ionization or shock heating.  Since a strong X-ray source
(the pulsar) is present in the system, it is essential to determine
whether the emission lines are primarily powered by ultraviolet
photons (as in typical symbiotic binaries) or X-rays, since the
ionization structure in the two cases will be quite different.  By
analogy with emission-line galaxies, we can use the location of the
dereddened line ratios $I$([\ion{O}{3}] $\lambda$5007)/$I$(H$\beta$)
and $I$([\ion{O}{1}] $\lambda$6300)/$I$(H$\alpha$) on a so-called BPT
diagram (Baldwin, Phillips, \& Terlevich 1981; Veilleux \& Osterbrock
1987) to probe the nature of the photoionizing continuum.
Emission-line galaxies known to be photoionized by ultraviolet
radiation from OB stars populate a distinct region of these diagrams
from those ionized by an X-ray continuum. The location of V2116 Oph on
such a diagram suggests that the nebula is ionized by thermal ultraviolet
radiation rather than the non-thermal X-ray power law continuum
expected from an accreting neutron star.

This being the case, we can apply the standard optical emission line
diagnostics used for estimating the physical conditions in symbiotic
nebulae.  Following Iijima (1981), we used the dereddened line
strengths of H$\beta$, \ion{He}{1} $\lambda$4471, and \ion{He}{2}
$\lambda$4686 to estimate that the hot component of this symbiotic
binary has a characteristic temperature $T_{\rm hot}\lesssim 90000$
K. (This estimate is an upper limit because the H$\beta$ line may be
depressed by optical depth effects; see Kenyon 1986 and the discussion
below.)  The ionizing source is much hotter than the photosphere of
the M giant but considerably cooler than the characteristic
temperature of the accreting neutron star.  We conclude that the
ionization source is an accretion disk around GX 1+4, which we discuss
further in Section 4.3. 

As noted by Davidsen et al. (1977), the large H$\alpha$/H$\beta$ ratio
indicates a high electron density and optical depth, leading to
significant Balmer line self-absorption effects. Our 1991 August
spectrum measures the Balmer decrement using the first six Balmer
lines, providing a significant constraint on the physical conditions
of the emission nebula.  Comparing the measured ratios to models of a
photoionized high-density slab of hydrogen (Drake \& Ulrich 1980), we
find they are 
consistent with electron temperature $T_e=2\times 10^4$ K,
electron density $N_e\approx 10^9$ cm$^{-3}$, optical depths
$\tau_{{\rm Ly}\alpha} \approx 10^6$ and $\tau_{{\rm H}\alpha}\approx
800$ in Ly$\alpha$ and H$\alpha$, and photoionization rate
\begin{equation}
R_{1C}\equiv 4\pi\int_{\nu_1}^{\infty} a_1(\nu)\frac{I_\nu}{h\nu}d\nu
	= 0.03 \mbox{\rm\ s}^{-1} ,
\end{equation}
where $I_\nu$ is the specific intensity of the external radiation
field, $a_1(\nu)$ is the bound-free cross-section of the hydrogen
ground state, $h=6.6\times 10^{-27}$ erg s is Planck's constant, and
$h\nu_1$=13.6 eV is the hydrogen Lyman edge energy.  The equivalent
Krolik \& McKee (1978) ionization parameter (the ratio of the ionizing
photon density to the electron density) is $\Gamma\equiv (4\pi/c
N_e)\int_{\nu_1}^\infty (I_\nu/h\nu)d\nu = 5\times 10^{-4}$,
where we have assumed an $8\times 10^4$ K blackbody radiation field.
The \ion{He}{1} line ratios are also completely inconsistent with case
B values. The large $I$(\ion{He}{1} $\lambda$6678)/$I$(\ion{He}{1}
$\lambda$5876)$\gtrsim 0.5$ ratio typically measured also implies a
dense, optically thick nebula (Proga, Miko{\l}ajewska, \& Kenyon
1994).

Our inference of high optical depth is confirmed by the great
strength of \ion{O}{1} $\lambda$8446 emission.  This line can be produced by
four mechanisms: recombination, collisional excitation, continuum
fluorescence, and Bowen fluorescence by \ion{H}{1} Ly$\beta$
(Grandi 1980; Rudy et al. 1990). The relative weakness of the \ion{O}{1}
$\lambda$7774 line rules out both recombination and collisional
excitation, while the extreme weakness of \ion{O}{1} $\lambda$7254 and
the absence of detectable \ion{O}{1} $\lambda$7990 rules out continuum
fluorescence (Grandi 1980).  The only remaining explanation,
Bowen fluorescence by Ly$\beta$, operates only in a gas which is
optically thick in H$\alpha$, yielding the relation 
\begin{equation}
I(\lambda8446)/I(\mbox{H}\alpha) = 1.8\times
10^{-5}/\epsilon_{\mbox{\tiny H}\alpha} ,
\end{equation}
where $\epsilon_{\mbox{\tiny H}\alpha}$ is the escape probability of an
H$\alpha$ photon and a solar O/H abundance ratio is assumed (Grandi 1980). Our
measured value of $I(\lambda8446)/I($H$\alpha)$=0.12 yields
$\epsilon_{\mbox{\tiny H}\alpha}=1.5\times 10^{-4}$, which corresponds
to an optical depth $\tau_{\mbox{\tiny H}\alpha}\sim 4500$
(Ferland \& Netzer 1979). This is even higher than the Balmer line estimate
above, possibly due to time variability in the optical depth of the
region. A prediction of the Bowen fluorescence explanation for the
strength of the $\lambda$8446 line is that the dereddened photon
number fluxes for the \ion{O}{1} $\lambda\lambda$1304, 11287 lines
should be identical to that of $\lambda$8446 (Rudy et al. 1990). 

We can use the 1991 August dereddened hydrogen and helium line
strengths to estimate the size and structure of the emission-line
nebula surrounding the binary.  For the H$^+$ region, using the
H$\beta$ emissivity $(4\pi j_{{\rm H}\beta}/N_p N_e) = 0.524\times 10^{-25}$
erg cm$^{-3}$ s$^{-1}$ appropriate for the conditions inferred above
from the measured Balmer decrement (Drake \& Ulrich 1980), we find
the Stromgren radius $R_{{\rm H}^+} = 3.7$ AU $D_{10}^{2/3}
N_9^{-2/3}$, where $D_{10}$ is the distance to the source in units of
10 kpc and $N_9$ is the electron density in units of $10^9$
cm$^{-3}$. The emissivity for the \ion{He}{1} $\lambda$5876 line is
$(4\pi j_{5876}/N_{{\rm He}^+} N_e) = 0.524\times 10^{-25}$ erg
cm$^{-3}$ s$^{-1}$ for $N_e=10^{-9}$ cm$^{-3}$ and a 10\% helium number
abundance, over a wide range of optical depth (Almog \& Netzer 1989),
from which we estimate $R_{{\rm He}^+} = 3.4$ AU $D_{10}^{2/3}
N_9^{-2/3}$. Finally, using the case B emissivity for \ion{He}{2}
$\lambda$4686 (Osterbrock 1989), we find $R_{{\rm He}^{++}}=1.3$ AU
$D_{10}^{2/3} N_9^{-2/3}$.

We note that the \ion{He}{1} ratios showed considerable variability
over time.  In particular, the two AAT spectra we acquired on 1993
September 10, exposed only 15 minutes apart, show an astonishing
difference: the \ion{He}{1} $\lambda\lambda$5876, 6678, 7065 and
\ion{Fe}{2} $\lambda\lambda$6433, 6516 lines, which are clearly
detected in the second AAT spectrum (and all of our spectra from other
epochs) are completely absent in the first AAT spectrum exposed only
15 minutes earlier! In addition, the \ion{He}{1} $\lambda$7281 line is
50\% weaker in the first spectrum. In all other respects, however, the
two spectra are essentially identical (see Figure 6 and Table 8).  The
variable \ion{Fe}{2} $\lambda\lambda$6433, 6516 lines are both from
the same multiplet (Moore 1972).  The rapid change sets an upper limit
of $\lesssim 1.8$ AU on the size of the He$^+$ region. This is
consistent with our calculated Stromgren radius only if $D\lesssim 4 N_9$
kpc.  We have no explanation for the rapid change in these line
strengths. It cannot be due to the 2-min rotational modulation
of the X-ray flux from the neutron star, since both exposures span at
least one pulse period.  In fact, the longer exposure (in which the
line strengths are so low) should have averaged over several pulse
periods, in which case we might have expected the lines to be stronger
in this exposure. 

All of our emission lines indicate a barycentric radial velocity
$V\approx -150$ km s$^{-1}$, consistent with previous measurements
(Glass \& Feast 1973; Davidsen et al. 1977).  Several of our
high-resolution H$\alpha$ spectra show interesting evolution of the
emission line profile (Figure 6). On 1994 Feb 25, the central
H$\alpha$ peak was collapsed, with a peak on the blue wing shifted
about 150 km s$^{-1}$ from the line center and an absorption
``notch'' blue-shifted about 90 km s$^{-1}$ from the line center (left
panel of Figure 6). A week later, both the blue-shifted peak and the
notch are still present, but the central peak is much stronger (center
panel of Figure 6). Four months later, no complex substructure is
detected in the line profile (right panel of Figure 6). The 1994 March
7 line profile is comparable to the S-2 type profile identified by van
Winckel et al. (1993) in their survey of symbiotic stars, while the
1994 July profile to the S-1 type. These variations may be associated
with cyclic changes over the binary orbit (Kenyon 1986). Sood et
al. (1995) have proposed to search for orbital-phase modulation of the
H$\alpha$ line profile in order to determine the binary period. The
observed line profile variations may be due to orbital-phase variation
of the X-ray illumination of the red giant's atmosphere (Schwank,
Schmutz, \& Nussbaumer 1997). 

\section{DISCUSSION}

\subsection{Multiwavelength correlations}

Previous authors have cautioned that the association of GX~1+4 and
V2116 Oph, although likely, was not completely secure (e.g., Verbunt,
Wijers, \& Burm 1990). Now, however, the excellent coincidence of the
new optical and X-ray astrometry, our evidence for X-ray heating of
V2116 Oph during the 1993 September GX~1+4 outburst, and the recent
discovery of 2-min optical pulsations from V2116 Oph (Jablonski et
al. 1997) all reaffirm the identification of V2116 Oph as the binary
companion of GX~1+4.  Correlated X-ray/optical intensity behavior is
common in LMXBs (e.g., van Paradijs \& McClintock 1995), and many
investigators have searched for such correlations in GX 1+4. 

Throughout the 1970s X-ray bright state of GX 1+4, extremely strong
H$\alpha$ emission was observed from V2116 Oph. During the 1983--1984
X-ray low state established by a series of {\em EXOSAT} non-detections
(Hall \& Davelaar 1983; Mukai 1988), H$\alpha$ emission was present on
three occasions but undetected on another (Whitelock et
al. 1983; Whitelock 1984; see also Table 4). An examination of Table 6
indicates that several forbidden species ([\ion{O}{3}], [\ion{Fe}{7}],
[\ion{O}{1}]) had substantially stronger line emission during the 1988
July optical observation (contemporaneous with the weak X-ray emission
and steady spin-down in GX 1+4; see Makishima et al. 1988). During
1991--1994 (when the X-ray flux was higher), these forbidden features
were again weak or even undetected, suggesting that the bright X-ray
states may be accompanied by higher electron densities in the
symbiotic nebula.

Greenhill et al. (1995) found photometric evidence for a correlated
increase in H$\alpha$ strength during the 1993 September hard X-ray
flare, with the optical flare onset leading the X-ray onset by about a
month but both ending at roughly the same time. Manchanda et
al. (1995) claimed a correlation between spectroscopic H$\alpha$
strength and hard X-ray flux during 1994 March--June. However, the
case for the Manchanda et al.  correlation is weak, both due to
undersampling of the optical history and because the optical data was
correlated against a BATSE pulsed count rate history which was
uncorrected for detector viewing angle or detector response.

More recently, an extended X-ray low state beginning in 1996 September
(Chakrabarty, Finger, \& Prince 1996; Cui \& Chakrabarty 1996) was
accompanied both by clearly detected H$\alpha$ emission on October 6
(M. J. Coe 1996, private communication) and an absence of detectable
H$\alpha$ emission on October 16 (Sood et al. 1996).  Overall,
however, the long-term photometric intensity history of both the
H$\alpha$ feature and the {\em VRI} bands appears to be correlated
with the hard X-ray intensity (Phillips, Greenhill, \& Hill
1996). Still, on short time scales, our data indicate that the X-ray
activity of the pulsar is an unreliable predictor of the optical and
infrared properties of the giant.  A comparison of our H$\alpha$ line
strength measurements in Tables 6--8 with simultaneous BATSE
measurements of the hard X-ray intensity in Table 1 do not show clear
evidence of a correlation.  Long-term simultaneous histories of X-ray
intensity and H$\alpha$ emission line strength, if both sampled on
sufficiently short time scales, may provide an indirect probe of the
accretion disk.

\subsection{Source distance and the evolutionary status of the mass donor}

Archival X-ray measurements place some constraints on the distance to
the system. Following the argument by Chakrabarty et al. (1993), the
steady pulsar spin-up rate of $\dot\nu\approx 6\times 10^{-12}$ Hz
s$^{-1}$ during the 1970s X-ray bright state (Nagase 1989 and references
therein) and its bolometric X-ray flux of $8\times 10^{-9}$ erg
cm$^{-2}$ s$^{-1}$ (Doty, Hoffman, \& Lewin 1981) set a limit of
$D\gtrsim 3$ kpc on the distance and $L_x\gtrsim 8\times 10^{36}$ erg
s$^{-1}$ on the X-ray luminosity.  We can infer an upper limit of
$D\lesssim 15$ kpc by requiring that GX 1+4 not exceed the Eddington
critical luminosity ($L_{\rm Edd}=2\times 10^{38}$ erg s$^{-1}$) for
spherical accretion onto a neutron star. 

The $K$-band bolometric correction for an M5 III star [for which
$(J-K)_0= 1.23$; Bessell \& Brett 1988] is BC$_K=3.10\pm 0.05$ (Frogel
\& Whitford 1987), which (with our measured 1993 June values of
$K=8.13\pm 0.02$ and $A_K=0.56\pm 0.08$) gives an apparent bolometric
magnitude of $m_{\rm bol}=10.67\pm 0.10$.  Combined with the effective
temperature $T_{\rm eff}=3470$ K corresponding to our M5 III spectral
classification (Dyck et al. 1996), we can deduce a distance-radius
relation for the red giant: $D=5.6 R_{100}$ kpc, where $R_{100}$ is
the radius of the red giant in units of 100 $R_\odot$.  Furthermore,
the red giant's luminosity must scale as $L_g = 3500 D^2_{10}
L_\odot = 1100 R^2_{100} L_\odot$. Similarly, the X-ray luminosity
must scale as $L_x = 3\times 10^{37} R^2_{100}$ erg s$^{-1}$,
roughly consistent with the relation derived from X-ray heating in
Section 3.1.  The $D\gtrsim 3$ kpc limit from X-ray timing thus
implies that $R_g\gtrsim 50 R_\odot$ and $L_g\gtrsim 280
L_\odot$.

We can use this to further constrain the nature of the companion and
the distance to the system. The spectral and luminosity classification
of V2116 Oph permits two possibilities for its evolutionary status
(see, e.g., Hansen \& Kawaler 1994). It may be a low mass star
(zero-age main sequence mass $M_{\rm ZAMS}\lesssim 2\,M_\odot$) on the
first-ascent red giant branch (FGB), where a degenerate helium core is
fed by hydrogen shell burning in the envelope.  Alternatively, it
may be a low or intermediate mass star ($0.8 M_\odot<M_{\rm
ZAMS}\lesssim 10 M_\odot$) beginning its ascent of the asymptotic
giant branch (AGB), where a degenerate carbon-oxygen core is fed by
helium and hydrogen shell burning. Statistically speaking, most stars
later than M5 III are on the AGB, while most earlier type giants are
on the FGB (Judge \& Stencel 1991).

The luminosity of both low-mass FGB and AGB stars is almost uniquely
determined by the core mass $m_c$, independent of the mass in
their outer envelopes (Joss, Rappaport, \& Lewis 1987 and references
therein). The maximum luminosity at the tip of the low-mass FGB ($M_{\rm bol}
\gtrsim -3.2$) occurs when the degenerate helium core grows large enough
($m_c\approx 0.45 M_\odot$) to ignite, leading to an explosive
runaway (helium flash) and sending the star onto the horizontal
branch.  For V2116 Oph, this FGB luminosity limit would imply
$D\lesssim 6$~kpc. However, our unheated M6 III classification 
corresponds to $T_{\rm eff}=3380\pm95$ K, unusually low for an FGB
star.  Stellar evolution calculations suggest that only a metal-rich
star with $M_{\rm ZAMS}\lesssim 2 M_\odot$ could have evolved to this
temperature before helium flash, and even so must be near the FGB tip
(Sweigart, Greggio, \& Renzini 1989). If V2116 Oph is on the FGB, it
must be close to helium flash, with the following properties: $0.8
M_\odot\lesssim M_{\rm ZAMS} \lesssim 2 M_\odot$, $280 L_\odot
\lesssim L_g \lesssim 1300 L_\odot$, $50 R_\odot \lesssim R_g\lesssim
110 R_\odot$, and 3 kpc $\lesssim D \lesssim $ 6 kpc. This 
implies an X-ray luminosity in the range $8\times 10^{36}$ erg s$^{-1}
\lesssim L_x\lesssim 3\times 10^{37}$ erg s$^{-1}$ for GX 1+4.

If, on the other hand, V2116 Oph is on the AGB, its spectral
classification and the red giant luminosity constraint placed by the
X-ray Eddington limit both suggest that it is just beginning it
ascent. Without question, the stability of the infrared magnitudes of
V2116 Oph (see Table 3) preclude it from being a long-period variable
(LPV), since these stars undergo regular $\gtrsim 1$ mag variations in
the infrared (Whitelock 1987). LPVs are generally highly evolved AGB
stars in which thermal pulsations have set in (Iben \& Renzini
1983). By contrast, the X-ray upper limit on the distance strongly
constrains V2116 Oph to lie near the bottom of the AGB luminosity
range. We can set a {\em minimum} AGB luminosity by noting that
most AGB stars have $m_c\gtrsim 0.6 M_\odot$, corresponding to
$M_{\rm bol}\lesssim -4.7$. Thus, if V2116 Oph is an AGB star, it
probably has the following properties: $0.8 M_\odot \lesssim M_{\rm
ZAMS} \lesssim 10 M_\odot$, $4800 L_\odot \lesssim L_g \lesssim
8000 L_\odot$, $210 R_\odot \lesssim R_g \lesssim 270 R_\odot$,
and 12 kpc $\lesssim D \lesssim $ 15 kpc.  This implies an X-ray
luminosity in the range $1\times 10^{38}$ erg s$^{-1} \lesssim L_x
\lesssim 2\times 10^{38}$ erg s$^{-1}$ for GX 1+4.

The two different possibilities for which giant branch V2116 Oph is
ascending lead to two distinct distance ranges, and it is not obvious
which possibility is more likely.  Unfortunately, there is no
straightforward way to observationally distinguish the FGB and AGB for
M giant stars (Judge \& Stencel 1991).  Statistically, an
intrinsically M6 III star is somewhat likelier to be on the AGB. In
the case of an FGB star, our lower luminosity limit $L_g\gtrsim
280 L_\odot$ implies a core mass $m_c\gtrsim 0.34 M_\odot$ (Joss
et al. 1987), which places the star close enough (within $\sim 10^7$
yr) to helium flash to make finding it there somewhat unlikely.
However, the measured interstellar extinction to V2116 Oph
($A_V\approx 5$) is rather low for the $\gtrsim 10$ kpc line of sight
through the Galactic plane which is required if the star is on the
AGB, while it is quite reasonable for a closer ($\lesssim 6$ kpc)
distance (e.g., Spitzer 1978).  Also, a distance of $\lesssim 4$ kpc
is apparently required to reconcile the Stromgren radius of the He$^+$
region with the size implied by the rapid line variability we observed
in 1993 September, which is inconsistent with the AGB solution. We
therefore consider it most likely that V2116 Oph is on the FGB at a
distance of 3--6 kpc. A summary of our conclusions is given in Table 9.

\subsection{The accretion disk}

Since optical emission line diagnostics suggest that the symbiotic nebula
enshrouding the binary is powered by ultraviolet radiation rather than
the neutron star's X-ray emission (Section 3.3), we conclude
that an accretion disk is present in the system.  The recent
discovery of 2-min coherent optical pulsations at short wavelengths
(Jablonski et al. 1997) is probably a sign of X-ray reprocessing in
this accretion disk (Chester 1979).  Our emission line observations
provide a crude measure of the shape of the disk's ultraviolet
continuum spectrum through the Iijima $K$-parameter, defined as the
number of He$^+$-ionizing photons relative to the H$^0$-ionizing and
He$^0$-ionizing photons. This parameter can be computed using optical
emission line strengths (Iijima 1981),
\begin{equation}
K = \frac{\int_{4\nu_1}^{\infty} (L_\nu/h\nu) d\nu}
         {\int_{\nu_1}^{4\nu_1} (L_\nu/h\nu) d\nu}
  = \frac{2.2\, I(\mbox{\rm \ion{He}{2}\ }\lambda4686)}
      {4.16\, I(\mbox{\rm H}\beta) + 
         9.94\, I(\mbox{\rm \ion{He}{1}\ }\lambda4471)},
\end{equation}
where $L_\nu$ is the overall disk luminosity at frequency $\nu$, and 
$h\nu_1=13.6$~eV. Our 1991 August line strengths yield $K=0.03$. We
note that this value is actually an upper limit, since H$\beta$
strength may be depressed by optical depth effects (Kenyon 1986). 

The overall luminosity from a geometrically thin, optically thick
accretion disk is given by
\begin{equation}
L_\nu \approx \frac{16\pi^2h\nu^3}{c^2}
        \int_{r_{\rm in}}^{r_{\rm out}}
        \frac{r\,dr}{\exp[h\nu/kT(r)] - 1} ,
\end{equation}
where $k=1.4\times 10^{-16}$ erg K$^{-1}$ is Boltzmann's constant, $c$
is the velocity of light, $r_{\rm in}$ and $r_{\rm out}$ are the inner
and outer radii of the disk, and $T(r)$ is the disk's temperature
profile as a function of mid-plane radius $r$.  For an X-ray heated
disk, we may consider the disk as consisting of up to three distinct
regions: an innermost region powered primarily by internal viscous
dissipation, a middle region of ``shallow'' X-ray heating, and an
outer region of ``deep'' X-ray heating (e.g., Cunningham 1976; Arons
\& King 1993).  We consider each of these regions in turn.

The standard temperature profile for an unirradiated thin accretion
disk is set by internal viscous dissipation (Shakura \& Sunyaev 1973;
Frank, King, \& Raine 1992)   
\begin{equation}
T_0 = \left(\frac{3GM_x\dot M}{8\pi\sigma r^3}\right)^{1/4} =
     1.3\times 10^4\ M_{1.4}^{1/4}\, \dot M_{-9}^{1/4}\, r_{10}^{-3/4} 
     {\rm\ \ K} , 
\end{equation}
where $M_{1.4}$ is the neutron star mass $M_x$ in units of 1.4
$M_\odot$, $\dot M_{-9}$ is the mass accretion rate $\dot M$ in units
of $10^{-9}\ M_\odot$~yr$^{-1}$, and $r_{10}$ is the mid-plane radius
in units of $10^{10}$ cm.  At sufficiently small radii, this internal
heating will dominate over X-ray heating in setting the disk
temperature.  Beyond a critical radius, however, X-ray heating of the
disk surface will modify the temperature profile according to
\begin{equation}
T_{\rm irr}=(T_0^4 + T_x^4)^{1/4} .
\end{equation}
If we assume that the disk is irradiated by a central X-ray point
source, then we can write $T_x$ as
\begin{equation}
T_x = \left[\frac{L_x(1-\eta_d)}{4\pi\sigma r^2}
   \cos\psi\right]^{1/4} \approx \left[\frac{L_x(1-\eta_d)}
   {4\pi\sigma r^2} \left(\frac{dH}{dr} - \frac{H}{r}\right)\right]^{1/4},
\label{eq-tx}
\end{equation}
where $\eta_d$ is the X-ray albedo of the disk, $\psi$ is the angle
between the normal to the disk surface and the vector from the neutron
star, and $H$ is the disk's scale height.   Thus, we see that the
temperature profile due to X-ray heating is sensitive to the
functional form of $H(r)$ through the approximation for $\cos \psi$
(which assumes $H\ll r$).

In the shallow X-ray heating region, the surface temperature profile is
set by X-ray heating but the {\em central} temperature of the disk is
still set by internal viscous dissipation.  Thus, the disk thickness
has the usual value for a standard unirradiated thin disk (Shakura \&
Sunyaev 1973; Frank et al. 1992),
\begin{equation}
H = 1.7 \times 10^8\ \alpha^{-1/10}\, M_{1.4}^{-3/8}\, \dot M_{-9}^{3/20}\,
     r_{10}^{9/8} {\rm\ cm} ,
\label{eq-h98}
\end{equation}
where $\alpha$ is an order unity dimensionless parameterization of the
kinematic viscosity $\nu_{\rm visc}=\alpha c_s H$, and $c_s$ is the 
isothermal sound speed.  We thus find
\begin{equation}
T_x = 2.3\times 10^4\ \alpha^{-1/40}\, (1-\eta_d)^{1/4}\, M_{1.4}^{5/32}\,
     \dot M_{-9}^{23/80}\, r_{10}^{-15/32}{\rm\ K} .
\end{equation}
Shallow X-ray heating will dominate internal viscous heating in
setting the disk surface temperature at radii exceeding roughly $10^9$
cm.  

If $T_x^4 \gtrsim \tau T_0^4$, then X-ray heating will also dominate
internal heating in setting the {\em central} temperature of the disk
(Lyutyi \& Sunyaev 1976; Spruit 1995).   Here, $\tau$ is the disk's
optical depth, given by $\tau= 54\ \alpha^{-4/5}\, \dot M_{-9}^{1/5}$
for a standard $\alpha$-disk (Frank et al. 1992).  In this
region of deep X-ray heating, the vertical structure of the disk
will not obey the $H\propto r^{9/8}$ relationship in equation
(\ref{eq-h98}), but will instead follow a $H\propto r^{9/7}$ power
law (e.g., Cunningham 1976; Vrtilek et al. 1990; Arons \& King 1993).
Deep X-ray heating ``puffs up'' the outer disk so that
\begin{equation}
H = 1.1\times 10^9\ (1- \eta_d)^{1/7}\, M_{1.4}^{-3/7}\, \dot M_{-9}^{1/7}\,
  r_{10}^{9/7} {\rm\ cm} ,
\label{eq-h97}
\end{equation}
from which we find 
\begin{equation}
T_x = 2.7\times 10^4\ (1-\eta_d)^{2/7}\, M_{1.4}^{1/7}\, \dot M_{-9}^{2/7}\,
   r_{10}^{-3/7}{\rm\ K} .
\end{equation}
Deep X-ray heating will dominate over shallow heating at radii
exceeding roughly $2\times 10^{10}$ cm. 

With the appropriate choice of temperature profile, equations (4) and
(5) can be used to predict the Iijima $K$-parameter that should be
measured due to photoionzation by the disk's continuum emission.  The
most important parameter for determining $K$ will be the inner disk
radius $r_{\rm in}$. For an X-ray pulsar, $r_{\rm in}$ should be
roughly equal to the magnetospheric radius $r_m$, where the
neutron star magnetosphere disrupts the Keplerian disk flow
(e.g. Frank et al. 1992)
\begin{equation}
r_{\rm in} = r_m = 3\times 10^{8} \mbox{\rm\ cm\ }
      \left(\frac{L_x}{10^{37}\mbox{\rm\ erg s$^{-1}$}}\right)^{-2/7}
      \left(\frac{B}{10^{12}\mbox{\rm\ G}}\right)^{4/7} ,
\label{eq-alfven}
\end{equation}
where $B$ is the surface dipole magnetic field of the neutron star. If
GX 1+4 has a magnetic field strength typical of X-ray pulsars ($B\sim
10^{12}$ G), then we expect $r_{\rm in}\approx 3\times 10^8$ cm.
In this case, there will be a hot inner region of the disk which is
primarily powered internal viscous dissipation. Most of the photons
capable of ionizing He$^+$ will come from this region. 

However, GX 1+4 has undergone several torque reversals between steady
spin-up and spin-down of the pulsar (Makishima et al. 1988;
Chakrabarty et al. 1997). Standard accretion torque theory (e.g.,
Ghosh \& Lamb 1979) would therefore predict that the pulsar is
spinning near its equilibrium spin period, which occurs when $r_m$
(and thus $r_{\rm in}$) is close to the corotation radius $r_{\rm
co}=(GM_xP_{\rm spin}^2/4\pi^2)^{1/3}$,where the magnetic field
lines move at the local Kepler velocity. For a slow rotator like GX
1+4, the corotation radius is very large ($r_{\rm co}\approx 3\times
10^9$ cm).   An ultrastrong pulsar magnetic field ($B\sim 10^{14}$
G) would be required to disrupt the disk at such a large distance from
the neutron star.  The innermost region of this disk would be
dominated by shallow X-ray heating, and would be much cooler than the
inner edge of a disk terminating at $3\times 10^8$ cm; the expected
\ion{He}{2} $\lambda$4686 line strength would be correspondingly
weaker. Since these two values of $r_{\rm in}$ will result in
significantly different temperature profiles for the disk, it is
interesting to compare the predicted values of $K$ in each case with
the value we measure. 

We have modeled the accretion disk in GX 1+4 as a standard thin disk
irradiated by a central X-ray source, as described above.  To provide
for a smooth transition between the different heating regimes in a
simple way, we have used the sum of equations (\ref{eq-h98}) and
(\ref{eq-h97}) for the disk thickness and calculated the temperature
profile according to equations (6), (7), and (8).  We assume
the disk cuts off sharply at an outer boundary set by the tidal radius
of the neutron star ($\approx 0.9 R_{\rm Roche}\sim 10^{13}$ cm; Frank
et al. 1992).  We consider two cases: that of an FGB donor and an X-ray
luminosity of $L_x\approx 10^{37}$ erg s$^{-1}$, and that of an AGB
donor and $L_x\approx 10^{38}$ erg s$^{-1}$.  In each case, we
consider a wide range of pulsar magnetic field strengths ranging from
$B\sim 10^{12}$ G (``weak'') to $B\sim 10^{14}$ G (``strong''), and
compute the inner disk radius according to equation (\ref{eq-alfven}).
Finally, in integrating the numerator of equation (4), we cut off the
disk spectrum above 1 keV, since the photoionization rate is already
negligible at this energy.  Our calculations are not sensitive to the
exact choice of either this upper bound or $r_{\rm out}$.  We assumed
that the disk had an effective X-ray albedo $\eta_d=0.9$ (de Jong, van
Paradijs, \& Augusteijn 1996). 

The resulting values for $K$ are plotted in Figure 7.  We see
immediately that given a measured $K$, the type of mass donor (and
thus the X-ray luminosity) strongly constrains the magnetic field
strength and thus the inner disk radius.  Specifically, if the mass
donor is an FGB star, then the pulsar {\em cannot} be near its
equilibrium spin period; instead, the accretion disk must terminate
well inside $r_{\rm co}$ in order to produce enough hard UV photons to
produce the observed \ion{He}{2} $\lambda$4686 strength.  In this
case, a surface dipole magnetic field strength of $\sim 5\times
10^{12}$ G is expected for the pulsar.  On the other hand, if the mass
donor is an AGB star, then the pulsar magnetic field {\em must} be
ultrastrong ($\sim 10^{14}$ G) in order to terminate the disk at a
relatively large radius; otherwise, a considerably stronger
\ion{He}{2} line ought to have been observed.  This would be
consistent with the pulsar spinning close to its equilibrium spin
period.  Although these results are based on a rather simple model for
the accretion disk and the photoionized nebula, our qualitative
conclusions should be quite secure.

Since we regard it as rather more likely that the mass donor is an FGB
star, this analysis raises important doubts about the applicability of
standard accretion torque theory in explaining the observed pulsar torque
reversals in GX 1+4. 

\subsection{Limits on the orbital period}

Given a neutron star mass $M_x=1.4\,M_\odot$ and a maximum mass
and minimum radius for the red giant, we can compute a lower limit on
the orbital period (independent of the mass transfer mode) by
requiring that the giant fit inside its Roche lobe. The size of the
Roche lobe is $(R_L/a) \approx 0.49 q^{2/3} [0.6 q^{2/3} + \ln(1
+ q^{1/3})]^{-1}$, where $R_L$ is the radius of a sphere with
the same volume as the giant's Roche lobe, $a$ is the binary
separation, and $q=M_g/M_x$ is the mass ratio of the red
giant and the neutron star (Eggleton 1983).  For the FGB case, we have
$M_g\lesssim 2 M_\odot$ and $R_g\gtrsim 50 R_\odot$,
implying that $P_{\rm orb}\gtrsim 100$ d. (Because of its low
effective temperature, V2116 Oph is likeliest to be very close to 
helium flash in the FGB case, where $R_g\approx 100
R_\odot$. This would imply $P_{\rm orb}\gtrsim 280$ d.) For the AGB
case, $M_g\lesssim 10 M_\odot$ and $R_g\gtrsim 210
R_\odot$, implying that $P_{\rm orb}\gtrsim 260$ d. Overall, we see
that the binary period for the system is probably of order a year or
more, by far the longest of any known LMXB. These conclusions are
included in Table 9.

If we assume that the mass transfer onto the neutron star is driven by
Roche lobe overflow, then we can use a similar argument to estimate an
upper limit on the orbital period by taking the minimum mass and
maximum radius of the giant and requiring that it fill its Roche
lobe. If we neglect mass loss, then $M_g\gtrsim 0.8 M_\odot$ for
both the FGB and AGB cases. In the FGB case, $R_g\lesssim 110
R_\odot$ and so $P_{\rm orb}\lesssim 480$ d, while in the AGB case
$R_g\lesssim 270 R_\odot$ and $P_{\rm orb}\lesssim 5$ yr. With
mass loss, we have $M_g\gtrsim 0.34 M_\odot$ for the FGB and
$M\gtrsim 0.6 M_\odot$ for the AGB, yielding $P_{\rm orb}\lesssim 2$
yr and $P_{\rm orb}\lesssim 6$ yr, respectively. However, we emphasize
that these upper limits on the size of the binary are only meaningful
if the giant fills its Roche lobe.

Due to the large variations in the accretion torque on GX 1+4 on
$\sim$100 d time scales, detection of orbital variations in the X-ray
pulse timing data has proven elusive (Chakrabarty et al. 1997).
Cutler, Dennis, \& Dolan (1986) suggested a 304 d binary period based
on an examination of the 1970s accretion torque history of GX 1+4, but
this remains unconfirmed and may be an artifact of the strong $1/f$
torque noise in the pulsar's spin behavior (Bildsten et al. 1997).
Sood et al. (1995) have proposed to measure the orbital period by
searching for cyclic variations in the H$\alpha$ emission line
profile, a promising technique given the strength of the line
feature. Another possible approach would be to look for a low
amplitude cycle in the infrared magnitudes due to orbit-modulated
X-ray heating. For heating by a steady X-ray source, the peak-to-peak
fractional change in the red giant's luminosity should be $(\Delta
L_g/L_g) \lesssim 1 - (T_{\rm cool}/T_{\rm hot})^4$, with
the exact amplitude depending upon the binary inclination
angle. Taking $T_{\rm cool}=3370$ K and $T_{\rm hot}=3470$ K, we
expect infrared variations of $\Delta K\lesssim 0.1$ mag. The scatter
in the current infrared photometric data is $\sim 0.2$ mag (see Table
3), but more precise measurements are feasible.  In addition, a
precise radial velocity curve for the red giant should be measurable
using the infrared CO absorption bands near 2.3 $\mu$m.  Since V2116
Oph is quite bright in this band ($K\approx 8$), it is an excellent
candidate for such measurements. 

\subsection{Mass loss from the red giant and binary mass transfer}

We first consider the possibility that mass donor fills its Roche
lobe.  A donor mass of $M_g\lesssim 1 M_\odot$ is required to
avoid dynamically unstable mass transfer, and even then the mass
transfer rate in a binary as wide as GX 1+4 must be highly
super-Eddington (Kalogera \& Webbink 1996), with a value of $\sim
10^{-7} M_\odot$~yr$^{-1}$ for an FGB star (Webbink et al. 1983) and
$\sim 10^{-6}$ for an AGB star (de Kool, van den Heuvel, \& Rappaport
1986).  Since Roche lobe overflow would thus imply prodigious mass loss
from the binary, an earlier suggestion of radio jets from
GX 1+4 (Manchanda 1993) is intriguing. However, more sensitive radio
observations indicate that the proposed jets are probably field
sources unassociated with GX 1+4 (Fender et al. 1997; Marti et
al. 1997).  Further searches for evidence of a strong outflow from GX
1+4 would be of great interest. 

Steady accretion in LMXBs with Roche-lobe-filling evolved donors
should be rare due to dwarf-nova-like instabilities in the accretion
disk (King, Kolb, \& Burderi 1996). A binary as wide as GX 1+4 will
undergo stable mass transfer only for the short time that the red
giant mass remains $\gtrsim 0.9 M_\odot$, and will spend the remainder
of its X-ray-emitting lifetime (i.e. until the red giant envelope is
exhausted or ejected) as a soft X-ray transient (King et al. 1997b).
Systems whose donor envelope is still sufficiently massive to avoid
the disk instability must still somehow shield the donor from X-ray
irradiation, possibly with an accretion disk corona, in order to avoid
violent thermally-unstable mass transfer (King et al. 1997a).  Since
GX 1+4 is a persistent X-ray source, must be in a very short-lived
($\lesssim 10^6$ yr) stable accretion state if powered by
Roche-lobe-overflow, and it will soon become a soft X-ray transient.

An alternative scenario is that the V2116 Oph does not fill its Roche
lobe, but that the accretion disk around GX 1+4 forms from the wind of
the red giant.  This scenario is particularly attractive in the
context of explaining the accretion torque behavior of the
pulsar. Chakrabarty et al. (1997) found that the torque on the pulsar
is anticorrelated with the hard X-ray luminosity (which is presumably
proportional to $\dot M$), the opposite of what is predicted by
standard accretion torque theory (Ghosh \& Lamb 1979). To explain
these observations, Nelson et al. (1997) have revived earlier
suggestions (Makishima et al. 1988; Dotani et al. 1989) that the
spin-down may be due to episodic formation of a retrograde accretion
disk, with disk material rotating with the opposite sense as the
pulsar.  Although such a scenario is difficult to understand if the
mass donor fills its Roche lobe, it seems plausible if the disk were
formed from the slow, dense wind of the red giant.  In this case, the
unusually strong surface magnetic field invoked by standard accretion
torque theory (see Section 4.3) is not required, since the pulsar need
not be near its equilibrium spin period.  This also fits nicely with
two other inferences we made previously: that the mass donor is
probably an FGB star (Section 4.2), and that the pulsar is probably
not near its equilibrium spin period if the donor is an FGB star
(Section 4.3).

However, it is not clear that wind accretion can provide a
sufficiently high $\dot M$ in this system.  In typical symbiotic
binaries, ionization of the red giant wind by the hot component
results in radio continuum emission due to thermal bremsstrahlung,
whose intensity can be related to the giant's mass loss rate $\dot
M_w$ (Seaquist \& Taylor 1990).  A marginal detection of radio
emission from the GX 1+4/V2116 Oph system ($0.06\pm 0.02$ mJy at 6 cm;
Marti et al. 1997) establishes the scale for the mass loss rate
through a stellar wind of
\begin{equation}
-\dot M_w\approx 1\times 10^{-7} M_\odot \mbox{\rm\,yr$^{-1}$} 
    \left(\frac{v_w}{\mbox{\rm 10\ km s$^{-1}$}}\right)
    \left(\frac{D}{\mbox{\rm 10\ kpc}}\right)^{3/2}
    \left(\frac{S_{\rm 6cm}}{\mbox{\rm 0.06\ mJy}}\right)^{3/4} ,
\end{equation}
where $v_w$ is the wind velocity and we have used the mass-loss
formula of Wright \& Barlow (1975).  Naively assuming a spherically
symmetric wind, this mass loss rate is roughly consistent with that
expected for $v_w=10$ km s$^{-1}$, given the inferred electron density
($N_e\sim 10^9$ cm$^{-3}$) and scale size ($R\sim 10^{13}$ cm) of the
photoionized nebula enshrouding the binary.  However, such a small
mass-loss rate cannot power a $10^{38}$ erg s$^{-1}$ X-ray source, and
an unusually high accretion efficiency would be required to power even
a $10^{37}$ erg s$^{-1}$ source for the $\sim 10$ km s$^{-1}$ wind
velocity typically expected from red giants.

On the other hand, most of what is known about M giant wind velocities
is based on observations of late AGB stars; the winds from FGB or
early AGB stars might be somewhat faster, allowing a higher mass loss
rate in Equation (10).  Indeed, there is preliminary evidence for a
$\approx 100$ km s$^{-1}$ outflow from V2116 Oph (Chakrabarty, van
Kerkwijk, \& Larkin 1997, in preparation).  Thus, wind accretion
remains a viable possibility, at least for a sub-Eddington X-ray
source.  Even if the pulsar is currently wind-fed, nuclear evolution
of the red giant may still eventually lead to Roche lobe overflow.

\acknowledgements{
This paper was based in part on observations at Palomar Observatory,
which is owned and operated by the California Institute of
Technology. D.C. thanks Tom Prince and Fiona Harrison for acting as
faculty sponsors for the Palomar and {\em ROSAT} investigations during
his graduate work at Caltech.  We are grateful to Lee Armus, Colin
Aspin, Dave Buckley, Robert Knop, Stuart Lumsden, and Jesper Storm for
generously obtaining service observations for us, and to Malcolm Coe,
Juan Fabregat, Jules Halpern, and Shri Kulkarni for sharing archival
data with us. We thank Mike Read for providing astrometric
measurements of the UK Schmidt plate, and Hannah Quaintrell, Luisa
Morales, and Lucie Green for assistance in reducing the optical
data.  It is a pleasure to acknowledge Lars Bildsten, Malcolm Coe,
Marshall Cohen, Rob Fender, Bob Hill, Scott Kenyon, Davy Kirkpatrick,
Rob Nelson, Gerry Neugebauer, Angela Putney, Saul Rappaport, Ian
Thompson, and John Wang for helpful discussions and advice.  An
anonymous referee made several useful suggestions.

This work was supported in part by NASA grant NAG 5-2956 under the
{\em ROSAT} Guest Observer program. D.C. was supported by a NASA GSRP
Graduate Fellowship at Caltech under grant NGT-51184 and by a NASA
Compton Postdoctoral Fellowship at MIT under grant NAG
5-3109. P.R. thanks the Nuffield Foundation for financial support for
himself and L. Green.  }



\begin{deluxetable}{lr}
\tablecaption{Pulsed Hard X-ray Intensity of GX 1+4}
\tablewidth{3in}
\tablehead{ 
\colhead{Date} & \colhead{20--60 keV Pulsed Flux\tablenotemark{a}} }
\startdata
1991 Jul 01 & $33\pm3$ \nl
1991 Aug 10 & $36\pm4$ \nl
1993 Apr 06 & $13\pm2$ \nl
1993 Jun 25 & $29\pm3$ \nl
1993 Jun 30 & $44\pm3$ \nl
1993 Jul 05 & $31\pm5$ \nl
1993 Sep 08 & $124\pm3$ \nl
1994 Feb 25 & $8\pm4$ \nl
1994 Mar 07 & $15\pm3$ \nl
1994 Jun 30 & $15\pm4$ \nl
1995 Mar 12 & $75\pm4$ \nl
1995 Aug 14 & $41\pm4$ \nl
1996 Mar 26 & $103\pm5$ \nl
\enddata
\tablenotetext{a}{In units of $10^{-11}$ erg cm$^{-2}$ s$^{-1}$.}
\end{deluxetable}

\def\deg{\hbox{$^\circ$}}
\def\hh{\hbox{$^{\rm h}$}}
\def\mm{\hbox{$^{\rm m}$}}
\def\am{\hbox{$^{\prime}$}}
\begin{deluxetable}{llllll}
\tablewidth{5.5in}
\tablecaption{Optical Astrometry and Photometry of the GX 1+4 
   Field\tablenotemark{a}}
\tablehead{ 
\colhead{Star\tablenotemark{b}} & \colhead{RA (J2000)} & 
  \colhead{Dec. (J2000)} & \colhead{$B$} & \colhead{$V$} & \colhead{$R$} }
\startdata
V2116 Oph & 17\hh\,32\mm\,02\fs06 & --24\deg\,44\am\,45\farcs5 &20.78(5) &
    18.40(3) & 15.99(1) \nl
NNE  & 17\hh\,32\mm\,02\fs12 & --24\deg\,44\am\,39\farcs5 & $>23$    &
    21.0(3)  & 19.54(8) \nl
2    & 17\hh\,32\mm\,03\fs59 & --24\deg\,44\am\,24\farcs6 & 20.62(5) &
    18.57(4) & 16.60(2) \nl 
3    & 17\hh\,32\mm\,03\fs13 & --24\deg\,44\am\,30\farcs2 & 22.6(2)  &
    19.39(7) & 17.48(2) \nl 
4    & 17\hh\,32\mm\,03\fs05 & --24\deg\,44\am\,40\farcs3 & 22.6(2)  &
    19.7(1)  & 18.24(4) \nl 
6    & 17\hh\,32\mm\,03\fs36 & --24\deg\,44\am\,53\farcs9 & 22.2(2)  &
    18.75(4) & 16.82(1) \nl 
9    & 17\hh\,32\mm\,03\fs76 & --24\deg\,45\am\,33\farcs8 & 14.76(1) &
    13.13(1) & $<12.8$ \nl
10   & 17\hh\,32\mm\,00\fs37 & --24\deg\,44\am\,06\farcs5 & 17.28(1) &
    15.66(1) & 15.05(1) \nl 
12   & 17\hh\,32\mm\,00\fs79 & --24\deg\,44\am\,46\farcs0 & 22.8(3)  &
    19.53(8) & 17.80(3) \nl 
\enddata
\tablenotetext{a}{Epoch for astrometry is 1976 May 27. Epoch for photometry is
1993 June 29. Coordinates are accurate to 1 arcsec in RA and Dec.}
\tablenotetext{b}{Numbered stars follow scheme of Doxsey et al. (1977), who
also referred to V2116 Oph as GF.}
\end{deluxetable}

\begin{deluxetable}{lllllllll}
\tablewidth{6.5in}
\tablecaption{Infrared Photometry of V2116 Oph}
\tablehead{ 
\colhead{Date} & \colhead{Telescope} & \colhead{$K$} &
  \colhead{$J-H$} & \colhead{$H-K$} & \colhead{$K-L$} &
  \colhead{$K-L'$} & \colhead{Refs.}}
\startdata
1973 May 26 & SAAO 1.9-m &8.07(5)&1.54(7)&0.75(7)&\nodata  &\nodata &  1 \nl
1973 Aug 29 & SAAO 1.9-m &8.04(5)&1.54(7)&0.70(7)&0.71(7) &\nodata &  2 \nl
1974 Jun 13 & SAAO 1.9-m &8.12(5)&\nodata&0.80(7)&0.63(7) &\nodata &  2 \nl
1974 Jun 26 & SAAO 1.9-m &8.13(5)&\nodata&\nodata&\nodata  &\nodata &  2 \nl
1976 May 22 & SAAO 1.9-m &8.03(5)&1.49(7)&0.76(7)&0.74(7) &\nodata &  2 \nl
1977 Jul 23 & SAAO 1.9-m &8.09(5)&1.49(7)&0.74(7)&0.64(7) &\nodata &  2 \nl
1978 Jun 20 & SAAO 1.9-m &7.95(5)&1.51(7)&0.76(7)&0.73(7) &\nodata &  2 \nl
1980 Mar 23 & SAAO 1.9-m &8.05(5)&\nodata&\nodata & \nodata &\nodata &  3 \nl
1993 Apr 6  & Palomar 5-m&8.17(3)& 1.57(4) & 0.64(4) & \nodata &\nodata &  4 \nl
1993 Jun 26 & SAAO 1.9-m &8.10(2)& 1.60(2) & 0.61(3) &0.48(4) &\nodata &  4 \nl
1993 Jun 29 & UKIRT 3.8-m &8.17(8)&1.51(3)&0.58(8) & \nodata & 0.37(10)& 4 \nl
1993 Jun 30 & UKIRT 3.8-m &8.13(2)&1.48(4)&0.59(4) & \nodata & 0.45(8)& 4 \nl
1993 Jul 1  & UKIRT 3.8-m &8.13(2)&1.41(4)&0.49(4) & \nodata & \nodata& 4 \nl
1993 Jul 6  & UKIRT 3.8-m &8.13(2)&1.52(7)&0.44(5) & \nodata & 0.42(10) & 4 \nl
1993 Sep 09 & UKIRT 3.8-m &8.06(3)&1.40(6)&0.61(5) &\nodata &\nodata & 4 \nl
1993 Sep 11 & Teide 1.5-m & 8.06 & 1.38 & 0.62 & \nodata  &\nodata &  4 \nl
1994 Jun 30 & SAAO 1.9-m & 8.16 & 1.56 & 0.62 & 0.50 &\nodata & 4 \nl
1995 Aug 16 & SAAO 1.9-m & 8.17(1) & 1.59(5)&0.63(3)& \nodata  &\nodata & 4 \nl
\enddata
\tablerefs{(1) Glass \& Feast 1973; (2) Glass 1979; (3) Glass 1993, personal 
communication; (4) This work}
\end{deluxetable}

\begin{deluxetable}{llllll}
\tablewidth{6.5in}
\tablecaption{Optical Spectroscopy of V2116 Oph}
\tablehead{ 
\colhead{Date} & \colhead{Telescope} & \colhead{$\lambda\lambda$} &
  \colhead{$\Delta\lambda$} & \colhead{Ref.} & \colhead{Remarks}}
\startdata
1974 Jul 14 & Lick 3-m   & 4600--6700 & 10 & 1  & Many lines, including [Fe VII] \nl
1975 Jun 13 & Lick 3-m   & 4800--7300 & 10 & 1  & Many lines, including [Fe VII] \nl
1975 Aug 12 & Lick 3-m   & 5800--8250 & 10 & 1  & Many lines, including [Fe VII] \nl
1976 Apr 10 & AAT 3.8-m  & 4000--7325 & 1.3& 2  & Rich emission line spectrum \nl
1976 Aug 20 & AAT 3.8-m  & 4000--7325 & 1.3& 2,3& Rich emission line spectrum \nl
1983 Aug 7  & SAAO 1.9-m & unspecified&\nodata & 4  & Strong H$\alpha$ \nl
1983 Nov 4  & SAAO 1.9-m & unspecified&\nodata &4&H$\alpha$ weak or absent \nl
1984 Apr 22 & AAT 3.8-m  & 5200--11000& 10 & 2  & Rich emission line spectrum \nl
1988 Mar 21 & ANU 2.3-m  & unspecified&\nodata & 5  & Strong H$\alpha$ \nl
1988 Jun 18,22& WHT 4.2-m& 5000--9700 & 9 & 6 & Many lines \nl
1988 Jul 8  & Lick 3-m   & 4460--7230 &2.8 & 7,8& Rich emission line spectrum \nl
1991 Jul 2  & ANU 2.3-m   &4000--8500& 1.6 & 9,10  & Rich emission line spectrum \nl
1991 Aug 8  & Palomar 5-m &3800--5490& 2   & 2  & Rich emission line spectrum \nl
1991 Aug 8  & Palomar 5-m &5580--6860 & 1.5 &2  & Rich emission line spectrum \nl
1991 Aug 9  & Palomar 5-m &3800--5490 & 2 &  2  & Rich emission line spectrum \nl
1991 Aug 9  & Palomar 5-m &5580--6860 & 1.5 &2  & Rich emission line spectrum \nl
1993 Jun 30 & Palomar 5-m &5200--9500 & 6 &  2  & Rich emission line spectrum \nl
1993 Jul 1  & WHT 4.2-m  & 6200--7000 & 1.3 & 2  & H$\alpha$ profile \nl
1993 Sep 10 & NTT 3.5-m  & 6000--9000 & 3.6 & 2  & Rich emission line spectrum \nl
1993 Sep 10 & AAT 3.8-m  & 5800--7200 & 1.6 &  2  & He I lines missing  \nl
1993 Sep 10 & AAT 3.8-m  & 5800--7200 & 1.6 & 2  & Rich emission line spectrum \nl
1994 Feb 25 & AAT 3.8-m  & 6300--6850 & 0.5 & 2 & H$\alpha$ profile \nl
1994 Mar 7  & SAAO 1.9-m & 6200--6900 & 0.5 & 2 & H$\alpha$ profile \nl
1994 Jul 1  & SAAO 1.9-m & 6200--6900 & 0.5 & 2 & H$\alpha$ profile \nl
1996 Oct 6  & SAAO 1.9-m & 6200--6900 & 0.5 & 2 & Strong H$\alpha$ emission \nl
1996 Oct 16&Mt Stromlo 1.9-m&unspecified&\nodata&11&H$\alpha$ emission absent \nl
1996 Oct 19 & Mt Stromlo 1.9-m& unspecified&\nodata& 11 &H$\alpha$ emission absent \nl
\enddata
\tablerefs{(1) Davidsen et al. 1977; (2) This work; (3) Whelan et
al. 1977; (4) Whitelock et al. 1983; (5) Dotani et al. 1989; (6)
Shahbaz et al. 1996; (7) Gotthelf et al. 1988; (8) Halpern 1994, personal
communication; (9) Sood et al. 1991;  (10) Sharma et al. 1993; (11) Sood 
et al. 1996} 
\end{deluxetable}

\begin{deluxetable}{lrrrll}
\tablewidth{5in}
\tablecaption{Spectral Classification of V2116 Oph}
\tablehead{ 
\colhead{Date} & \colhead{[VO]} & \colhead{$I(7450)$} & \colhead{$S(7890)$} &
    \colhead{ST\tablenotemark{a}} & \colhead{ST\tablenotemark{b}} }
\startdata
1975 Aug 12 & 0.22 & 0.01 & 0.63 & M4.2 & M5 \nl
1984 Apr 22 & 0.28 & 0.00 & 0.69 & M4.8 & M5.5 \nl
1993 Jun 30 & 0.35 & 0.03 & 0.84 & M5.4 & M6 \nl
1993 Sep 10 & 0.12 & $-0.31$ & 0.30 & M3 & M3 \nl
\enddata
\tablenotetext{a}{Spectral type from Kenyon \& Fernandez-Castro (1987).}
\tablenotetext{b}{Spectral type from Terndrup et al. (1990).}
\end{deluxetable}

\begin{deluxetable}{llrrrrrrrrrr}
\tablewidth{6.9in}
\tablecaption{Selected Emission Line Strengths of V2116 Oph}
\tablehead{ 
  &  &  \multicolumn{2}{c}{1974/1975\tablenotemark{b}} &\multicolumn{2}{c}{1976 Apr 10} &
  \multicolumn{2}{c}{1976 Aug 20} &\multicolumn{2}{c}{1988 Jul 8\tablenotemark{c}} & 
  \multicolumn{2}{c}{1993 Jun 30} \\ 
\cline{3-4}\cline{5-6}\cline{7-8}\cline{9-10}\cline{11-12}
\colhead{$\lambda$\tablenotemark{a}} & \colhead{Ion} &
  \colhead{$F$\tablenotemark{d}} & \colhead{$I$\tablenotemark{e}} &
  \colhead{$F$\tablenotemark{d}} & \colhead{$I$\tablenotemark{e}} &
  \colhead{$F$\tablenotemark{d}} & \colhead{$I$\tablenotemark{e}} &
  \colhead{$F$\tablenotemark{d}} & \colhead{$I$\tablenotemark{e}} &
  \colhead{$F$\tablenotemark{d}} & \colhead{$I$\tablenotemark{e}} }
\startdata
4861 & \ion{H}{1} (H$\beta$)& 2.5  &5.9  & 2.6 & 6.1 & 2.6 & 6.2 & 1.2 & 2.9 & \nodata & \nodata\nl
4959 & [\ion{O}{3}]  & 0.8  & 1.6 & 0.5 & 1.0 & 0.6 & 1.2 & $<$2 & $<4$ &  \nodata & \nodata \nl
5007 & [\ion{O}{3}]  & 1.2  &2.3  & 1.5 & 2.9 & 1.1 & 2.1 & 4.8 & 9.3 & \nodata & \nodata\nl
5876 & \ion{He}{1} & 3.8  &2.8  &  1.7 & 1.2 & 1.5 & 1.1 & 2.1 & 1.5 & \nodata & \nodata\nl
6087 & [\ion{Fe}{7}] & 1.5  & 0.9 & $<$0.3 & $<$0.2 & 0.5 & 0.3 & 3.3 & 2.0 & \nodata & \nodata\nl
6300 & [\ion{O}{1}] & 0.4  &0.2  & 0.9 & 0.4 & \nodata & \nodata & 8.6 & 4.2 & \nodata & \nodata\nl
6563 & \ion{H}{1} (H$\alpha$) & 310  &121  & 110 & 42 & 72 & 28 & 240 & 94 & 254 & 99 \nl
6678 & \ion{He}{1} & 5.0  &1.8  & 2.2 & 0.8 & 0.9 & 0.3 & 3.3 & 1.2 & 4.0 & 1.4\nl
6830 & \ion{O}{6} Raman & $\approx$2  &$\approx$0.6  & 0.4 & 0.1 & 0.7 & 0.2 & 4.2 & 1.3 & \nodata & \nodata\nl
7065 & \ion{He}{1} & 8.3  &2.2  & 1.9 & 0.5 & 2.9 & 0.8 & 7.5 & 2.0 & 6.0 & 1.6\nl
7281 & \ion{He}{1} & \nodata  & \nodata & 1.7 & 0.4 & \nodata & \nodata & \nodata & \nodata & $<$1 & $<$0.2\nl
7774 & \ion{O}{1} & 5.5   &0.9  & \nodata & \nodata & \nodata & \nodata & \nodata & \nodata & 4.6 & 0.8\nl
8446 & \ion{O}{1} & \nodata & \nodata  & \nodata & \nodata & \nodata & \nodata & \nodata & \nodata & 58 & 6.9\nl
8498 & \ion{Ca}{2} & \nodata & \nodata  & \nodata & \nodata & \nodata & \nodata & \nodata & \nodata & 6.5 & 0.8\nl
8542 & \ion{Ca}{2} & \nodata & \nodata  & \nodata & \nodata & \nodata & \nodata & \nodata & \nodata & 11 & 1.3\nl
8662 & \ion{Ca}{2} & \nodata & \nodata  & \nodata & \nodata & \nodata & \nodata & \nodata & \nodata & 14 & 1.6\nl
\enddata
\tablenotetext{a}{Wavelength (\AA)}
\tablenotetext{b}{Average of three spectra (\cite{Davidsen77})}
\tablenotetext{c}{From Gotthelf et al. (1988) and Halpern (1994)}
\tablenotetext{d}{Observed flux in units of $10^{-15}$ erg cm$^{-2}$ s$^{-1}$}
\tablenotetext{e}{Dereddened flux in units of $10^{-13}$ erg cm$^{-2}$ s$^{-1}$, assuming $A_V=3.1\,E(B-V)=5$}
\end{deluxetable}

{\small
\begin{deluxetable}{llrr|llrr}
\tablewidth{6.9in}
\tablecaption{1991 August 8 Emission Line Strengths}
\tablehead{
\colhead{$\lambda$\tablenotemark{a}} & \colhead{Ion} &
  \colhead{$F$\tablenotemark{b}} & \multicolumn{1}{c|}{$I$\tablenotemark{c}} &
  \colhead{$\lambda$\tablenotemark{a}} & \colhead{Ion} &
  \colhead{$F$\tablenotemark{b}} & \colhead{$I$\tablenotemark{c}}}
\startdata
 3889& \ion{H}{1} (H8)     &  0.53   &  4.64  & 5411& \ion{He}{2}         & 0.52    & 0.62 \nl
 3970& \ion{H}{1} (H$\epsilon$)   &  0.65   &  5.16& 5425& \ion{Fe}{2} (49)    & 0.99    & 1.17 \nl
 4102& \ion{H}{1} (H$\delta$)   &  0.89   &  5.90& 5755& [\ion{N}{2}]        & 0.78    & 0.64  \nl
 4341& \ion{H}{1} (H$\gamma$)   &  1.94   &  9.22& 5876& \ion{He}{1}         & 12.81   &  9.30 \nl
 4363& [\ion{O}{3}] ?      &  0.89   &  4.11 & 5890& \ion{Na}{1}         &  0.76   &  0.54 \nl
 4388& \ion{He}{1} ?       &  0.45   &  2.02 & 5895& \ion{Na}{1}         &  0.62   &  0.45 \nl
 4416& [\ion{Fe}{2}] ?     & 0.57    & 2.46 & 5958& \ion{Si}{2}         & 0.35    & 0.23  \nl
 4471& \ion{He}{1}         &  0.32   &  1.27& 5979& \ion{Si}{2}         & 0.47    & 0.31  \nl
 4491& \ion{Fe}{2} (37)    & 0.41    & 1.59 & 5991& \ion{Fe}{2} (46)    & 0.97    & 0.63  \nl
 4555& \ion{Fe}{2} ?       & 0.58    & 2.04 & 6084& \ion{Fe}{2} (46)    & 0.54    & 0.32  \nl
 4583& \ion{Fe}{2} (37)    & 0.76    & 2.58 & 6103& \ion{Ca}{1} ?       &  0.25   &  0.14 \nl
 4640& \ion{C}{3}/\ion{N}{3}& 0.91    & 2.92 & 6148& \ion{Fe}{2} (74)    & 1.20    & 0.67  \nl
 4686& \ion{He}{2}     &    0.72 &    2.14 & 6156& \ion{O}{1} ?        &  0.31   &  0.17 \nl
 4713& \ion{He}{1}         &  0.43   &  1.23 & 6238& \ion{Fe}{2} (74)    & 1.28    & 0.66  \nl
 4861& \ion{H}{1} (H$\beta$)   & 13.45   & 31.62& 6248& \ion{Fe}{2} (74)    & 0.76    & 0.39  \nl
 4922& \ion{He}{1} ?       &  1.53   &  3.32 & 6300& [\ion{O}{1}]        &  1.76   &  0.86 \nl
 4959& [\ion{O}{3}]        &0.37     &0.76 & 6318& \ion{Fe}{2}         & 1.29    & 0.62 \nl
 4994& \ion{N}{2} ?        & 0.41    & 0.80 & 6347& \ion{Si}{2}         & 1.81    & 0.85 \nl
 5007& [\ion{O}{3}]    &    1.81 &    3.53& 6364& [\ion{O}{1}]        &  1.02   &  0.47\nl
 5016& \ion{He}{1} ?       &  2.14   &  4.11& 6369& \ion{Fe}{2} (40)    & 2.10    & 0.96 \nl
 5042& ?                   & 0.76    & 1.42 & 6384& \ion{Fe}{2}         & 1.21    & 0.55 \nl
 5056& \ion{Si}{2} ?    &    0.49 &    0.90& 6407& \ion{Fe}{2} (74)    & 0.11    & 0.05 \nl
 5133& \ion{Fe}{2} (35) &    0.26 &    0.43& 6417& \ion{Fe}{2} (74)    & 0.78    & 0.34  \nl
 5158& [\ion{Fe}{2}] (18)    & 0.95    & 1.53 & 6433& \ion{Fe}{2} (40)    & 1.95    & 0.85  \nl
 5169& \ion{Fe}{2} (42)    & 1.39    & 2.22 & 6443& \ion{Fe}{2}         & 0.60    & 0.26  \nl
 5184& \ion{Mg}{1}         &  0.26   &  0.41& 6456& \ion{Fe}{2} (74)    & 2.19    & 0.93  \nl
 5197& \ion{Fe}{2} (49)    & 0.58    & 0.90 & 6482& \ion{N}{2}          & 0.54    & 0.22  \nl
 5235& \ion{Fe}{2} (49)    & 0.66    & 0.97 & 6491& \ion{Fe}{2}         & 1.05    & 0.43  \nl
 5276& ?                   & 1.88    & 2.63 & 6516& \ion{Fe}{2} (40)    & 4.23    & 1.71  \nl
 5317& \ion{Fe}{2} (49)    & 1.40    & 1.87 & 6563& \ion{H}{1} (H$\alpha$) &487.18   &189.77 \nl
 5363& \ion{Fe}{2} (48)    & 0.80    & 1.01 & 6678& \ion{He}{1}         &14.05    & 4.99 \nl
 5376& [\ion{Fe}{2}] (19)   & 0.19    & 0.24& 6830& \ion{O}{6} Raman    & 0.51    & \hfill 0.16 \nl
\enddata
\tablenotetext{a}{Wavelength (\AA)}
\tablenotetext{b}{Observed flux in units of $10^{-15}$ erg cm$^{-2}$ s$^{-1}$}
\tablenotetext{c}{Dereddened flux in units of $10^{-13}$ erg cm$^{-2}$ s$^{-1}$, assuming $A_V=3.1\,E(B-V)=5$}
\end{deluxetable}
}

\begin{deluxetable}{llrrrrrr}
\tablewidth{6.9in}
\tablecaption{1993 September 10 Emission Line Strengths}
\tablehead{
  &  &  \multicolumn{2}{c}{NTT (0133 UT)} &\multicolumn{2}{c}{AAT (0904 UT)} &\multicolumn{2}{c}{AAT (0916 UT)} \\
\cline{3-4}\cline{5-6}\cline{7-8}
\colhead{$\lambda$\tablenotemark{a}} & \colhead{Ion} &
  \colhead{$F$\tablenotemark{b}} & \colhead{$I$\tablenotemark{c}} &
  \colhead{$F$\tablenotemark{b}} & \colhead{$I$\tablenotemark{c}} &
  \colhead{$F$\tablenotemark{b}} & \colhead{$I$\tablenotemark{c}} }
\startdata
 5755& [\ion{N}{2}]        & \nodata  &  \nodata &     0.12  &   0.10 & \nodata  &\nodata    \nl
 5876& \ion{He}{1}         & \nodata  & \nodata  &    0.73   &  0.53  &   5.89  &   4.27\nl
 5890& \ion{Na}{1}         & \nodata  & \nodata  &     0.41  &   0.29 &    0.66 &    0.47\nl
 5895& \ion{Na}{1}         & \nodata  & \nodata  &     0.21  &   0.15 &    0.31 &    0.22\nl
 5958& \ion{Si}{2}         & \nodata  & \nodata  &    0.19   &  0.12  &   0.13  &   0.09\nl
 5979& \ion{Si}{2}         & \nodata  & \nodata  &    0.14   &  0.09  &   $<$0.1& $<$0.07    \nl
 5991& \ion{Fe}{2} (46)    & \nodata  & \nodata  &    0.28   &  0.18  &   0.39  &   0.26\nl
 6084& \ion{Fe}{2} (46)    & \nodata  & \nodata  &    0.23   &  0.14  &   0.32  &   0.19\nl
 6103& \ion{Ca}{1} ?       & \nodata  & \nodata  &     0.06  &   0.04 &  $<$0.1  & $<$0.06    \nl
 6148& \ion{Fe}{2} (74)    & \nodata  & \nodata  &    0.36   &  0.20  &   0.53  &   0.29\nl
 6156& \ion{O}{1} ?        & \nodata  & \nodata  &     0.12  &   0.06 &  $<$0.1   & $<$0.06    \nl
 6238& \ion{Fe}{2} (74)    & \nodata  & \nodata  &    0.28   &  0.14  &   0.50  &   0.26\nl
 6248& \ion{Fe}{2} (74)    & \nodata  & \nodata   &    0.47   &  0.24  &   0.52  &   0.26\nl
 6300& [\ion{O}{1}]        & \nodata  & \nodata  &     $>$0.59 &   $>$0.29&    0.74 &    0.36\nl
 6318& \ion{Fe}{2}         & \nodata  & \nodata  &    0.35   &  0.17  &   0.50  &   0.24\nl
 6347& \ion{Si}{2}         & \nodata  & \nodata  &    0.60   &  0.28  &   0.74  &   0.35\nl
 6369& \ion{Fe}{2} (40)    &  1.41  &   0.65 &    0.59   &  0.27  &   1.16  &   0.534\nl
 6384& \ion{Fe}{2}         & $<$1.0 &  $<$0.45   &    0.38   &  0.17  &   0.60  &   0.27\nl
 6417& \ion{Fe}{2} (74)    & $<$1.0  & $<$0.45   &    0.23   &  0.10  &   0.38  &   0.17\nl
 6433& \ion{Fe}{2} (40)    &   0.72  &  0.32 &    $<$0.01   &  $<$0.004  &   0.89  &   0.39\nl
 6443& \ion{Fe}{2}         & $<$1.0  & $<$0.4  &    0.05   &  0.02  &   0.17  &   0.07\nl
 6456& \ion{Fe}{2} (74)    &   0.78  &   0.33 &  0.61 & 0.26    &   0.90  &   0.38\nl
 6482& \ion{N}{2}          & $<$1.0  & $<$0.4  &    0.66   &  0.28  &   0.27  &   0.11\nl
 6491& \ion{Fe}{2}         & $<$1.0  & $<$0.4  &    0.24   &  0.10  &   0.42  &   0.17\nl
 6516& \ion{Fe}{2} (40)    & $<$2.0     &$<$0.3  &  $<$0.01    & $<$0.004   &   1.40  &   0.57\nl
 6563& \ion{H}{1} (H$\alpha$)&244.62 &   95.29&   $>$183   & $>$71   &  310.96 &  121.13\nl
 6678& \ion{He}{1}         &   7.66  &   2.72 &    0.53   &  0.19  &   6.48  &   2.30\nl
 7002& \ion{O}{1} ?        & $<$1.0 & $<$0.3  &     0.24  &   0.07 &   $<$0.3& $<$0.08 \nl
 7065& \ion{He}{1}         &   16.85 &    4.49&     0.29  &   0.08 &   12.87 &    3.43\nl
 7115& \ion{C}{2} ?        &  $<$1.0 & $<$0.3  &    0.47   &  0.12  &   0.47  &   0.12\nl
 7136& [\ion{Ar}{3}] ?     &  $<$1.0 & $<$0.3  &   0.40    & 0.10   &  0.40   &  0.10\nl
 7155& [\ion{Fe}{2}] ?     &  $<$1.0 & $<$0.3  &    0.69   &  0.17  &   0.86 & 0.22 \nl
 {7172}&[\ion{Fe}{2}] ?    &$<$1.0 & $<$0.3  &   0.39    & 0.10 &   0.34  &  0.08 \nl
 {7222}& \ion{Fe}{2} (73)  & $<$1.0 & $<$0.2  &   0.44   &  0.11  &   0.54 & 0.13\nl
 7231& \ion{C}{2} ?        & $<$1.0 & $<$0.2  &   0.84   &  0.20  &   1.19  &   0.28\nl
 7281& \ion{He}{1}         &    3.30 &    0.76&    1.31  &   0.30 &    2.03 &    0.47\nl
 7291& [\ion{Ca}{2}]       & $<$1.0 & $<$0.2  &    0.57  &   0.13 &    0.61 &    0.14\nl
 7308& \ion{Fe}{2} (73)    & $<$1.0 & $<$0.2 &    0.37  &   0.08 & \nodata &\nodata \nl
 7712& \ion{Fe}{2} (73)    &   1.39  &   0.25 &  \nodata  &  \nodata & \nodata & \nodata \nl
 7774& \ion{O}{1}          &   6.69  &   1.14 &    \nodata  &  \nodata & \nodata & \nodata \nl
 8359& \ion{H}{1} (Pa22)   &    3.19 &    0.40 &   \nodata  &  \nodata & \nodata & \nodata \nl
 8374& \ion{H}{1} (Pa21)   &    3.09 &    0.39 &   \nodata  &  \nodata & \nodata & \nodata \nl
 8392& \ion{H}{1} (Pa20)   &    3.61 &    0.45 &   \nodata  &  \nodata & \nodata & \nodata \nl
 8413& \ion{H}{1} (Pa19)   &    5.41 &    0.66 &   \nodata  &  \nodata & \nodata & \nodata \nl
 8446& \ion{O}{1}          &   97.24 &   11.73 &   \nodata  &  \nodata & \nodata & \nodata \nl
 8467& \ion{H}{1} (Pa17)   &    5.19 &    0.62 &   \nodata  &  \nodata & \nodata & \nodata \nl
 8498& \ion{Ca}{2}         &  31.72  &   3.74 &    \nodata  &  \nodata & \nodata & \nodata \nl
 8542& \ion{Ca}{2}         &  42.28  &   4.88 &    \nodata  &  \nodata & \nodata & \nodata \nl
 8598& \ion{H}{1} (Pa14)   &    3.35 &    0.38 &   \nodata  &  \nodata & \nodata & \nodata \nl
 8662& \ion{Ca}{2}         &  30.49  &   3.34 &    \nodata  &  \nodata & \nodata & \nodata \nl
 8750& \ion{H}{1} (Pa12)   &    4.95 &    0.52 &   \nodata  &  \nodata & \nodata & \nodata \nl
\enddata
\tablenotetext{a}{Wavelength (\AA)}
\tablenotetext{b}{Observed flux in units of $10^{-15}$ erg cm$^{-2}$ s$^{-1}$}
\tablenotetext{c}{Dereddened flux in units of $10^{-13}$ erg cm$^{-2}$ s$^{-1}$, assuming $A_V=3.1\,E(B-V)=5$}
\end{deluxetable}

{\small
\begin{deluxetable}{lcc}
\tablewidth{5in}
\tablecaption{Inferred System Properties for GX 1+4/V2116 Oph}
\tablehead{ 
\colhead{Quantity} & \colhead{FGB mass donor} & \colhead{AGB mass donor} 
}
\startdata
ZAMS donor mass ($M_\odot$)& 0.8--2 & 0.8--10 \nl
Donor luminosity ($L_\odot$)& 280--1300 & 4800--8000 \nl
Donor radius ($R_\odot$)& 50-110 & 210--270 \nl
Distance (kpc) & 3--6 & 12--15 \nl
X-ray luminosity ($10^{38}$ erg s$^{-1}$)& 0.08--0.3 & 1--2 \nl
Pulsar surface dipole field (G) & $\sim 10^{12}$& $\sim 10^{14}$\nl
Inner disk radius (cm) & $\sim 10^8$ &$\sim 10^9$ \nl
Orbital period (d)& $\gtrsim 100$& $\gtrsim 260$\nl
\enddata
\end{deluxetable}
}


\clearpage
\centerline{FIGURE CAPTIONS}

\bigskip\noindent FIG. 1.---
$R$-band image of the GX 1+4 field, taken on 1993 September
10. The 10.4-arcsec-radius (90\%-confidence) {\em ROSAT} error circle
for GX 1+4 is also shown. In addition to V2116 Oph ($R=16$), a very
faint star (NNE, $R=19.5$) is also visible in the error circle.

\bigskip\noindent FIG. 2.---
Optical/near-infrared classification spectra of V2116
Oph. The top panel shows a low-resolution taken on 1993 June 30 from
the Palomar 5-m telescope. This spectrum corresponds to spectral type
M5.7 III. Due to the wide slit used, most of emission line features
are smeared out.  The bottom panel shows a higher resolution spectrum
taken on 1993 September 3 from the ESO 3.6-m NTT during a bright X-ray
outburst. This spectrum corresponds to M3 III.

\bigskip\noindent FIG. 3.---
Detailed optical emission line spectrum of V2116 Oph, taken
on 1991 August 9 from the Palomar 5-m telescope.

\bigskip\noindent FIG. 4.---
Low-resolution infrared spectrum of V2116 Oph in the $J$,
$H$, and $K$ bands, taken on 1993 April 6 from the Palomar 5-m
telescope.  The squares shows the corresponding IR photometry points.
Also shown is the 1993 June optical spectrum from the top panel of
Figure 2. For comparison, the dotted lines shows the best-fit reddened
($A_V=5$) blackbody curve, corresponding to a color temperature of
2590 K.  The blackbody curve is a poor fit to the M giant continuum due to
complicated infrared molecular opacities in cool stars. 

\bigskip\noindent FIG. 5.---
Spectacular rapid line variability in V2116 Oph on 1993 September 10,
during a bright X-ray flare. The top panel shows a 900 s exposure
centered at 0856 UT. The bottom panel shows a 150 s exposure centered
at 0915 UT. The following line features are mostly or completely absent
in the first spectrum and clearly present in the second: \ion{He}{1}
$\lambda\lambda$5876, 6678, 7065, and \ion{Fe}{2}
$\lambda\lambda$6433, 6516.   These lines are indicated by tick marks along
the bottom of both panels. The \ion{He}{1} $\lambda$7281 line
is about 50\% weaker in the first spectrum. The H$\alpha$ line was
saturated in the first spectrum and so could not be accurately
measured. All other line strengths are virtually identical in the two
spectra.  The bottom panel is representative of the spectrum typically
observed.

\bigskip\noindent FIG. 6.---
High-resolution H$\alpha$ spectra from three dates in
1994. The first two spectra were acquired about a week apart, and the
third spectrum about 4 months later. The center for all three lines
has a velocity around $-150$ km s$^{-1}$. A blue wing peak shifted
about 150 km s$^{-1}$ from the center is evident in the first two
panels, as is an absorption ``notch'' blue-shifted about 90 km
s${-1}$ from the center.

\bigskip\noindent FIG. 7.---
The predicted value for the Iijima $K$-parameter, the optical emission
line diagnostic defined in equation (4), as a function of the inner
radius of the accretion disk.  The curves shown were computed assuming
that the photoionization source is an X-ray heated accretion disk with
$r_{\rm out}=10^{13}$ cm and $\eta_d=0.9$.  The solid curve on the
left is for $L_x=10^{37}$ erg s$^{-1}$, as appropriate if the mass
donor is an FGB star at 3--6 kpc. The solid curve on the right is for
$L_x=10^{38}$ erg s$^{-1}$, as appropriate if the mass donor is an AGB
star at 12--15 kpc.  For both curves, various pulsar magnetic field
strengths are indicated by tick marks; the corresponding inner disk
radius is computed using equation (13). The dotted line indicates the
observed value $K=0.03$.  With our simpledisk model, the best
solutions consistent with the observed $K$ is $B\approx 5\times
10^{12}$ G for the FGB case, and $B\approx 7\times 10^{13}$ G for the
AGB case. 

\pagebreak
\pagestyle{empty}
\thispagestyle{empty}
\begin{figure}
\vspace{-2in}
\centerline{Figure 1}
\vspace{1in}
\centerline{\psfig{file=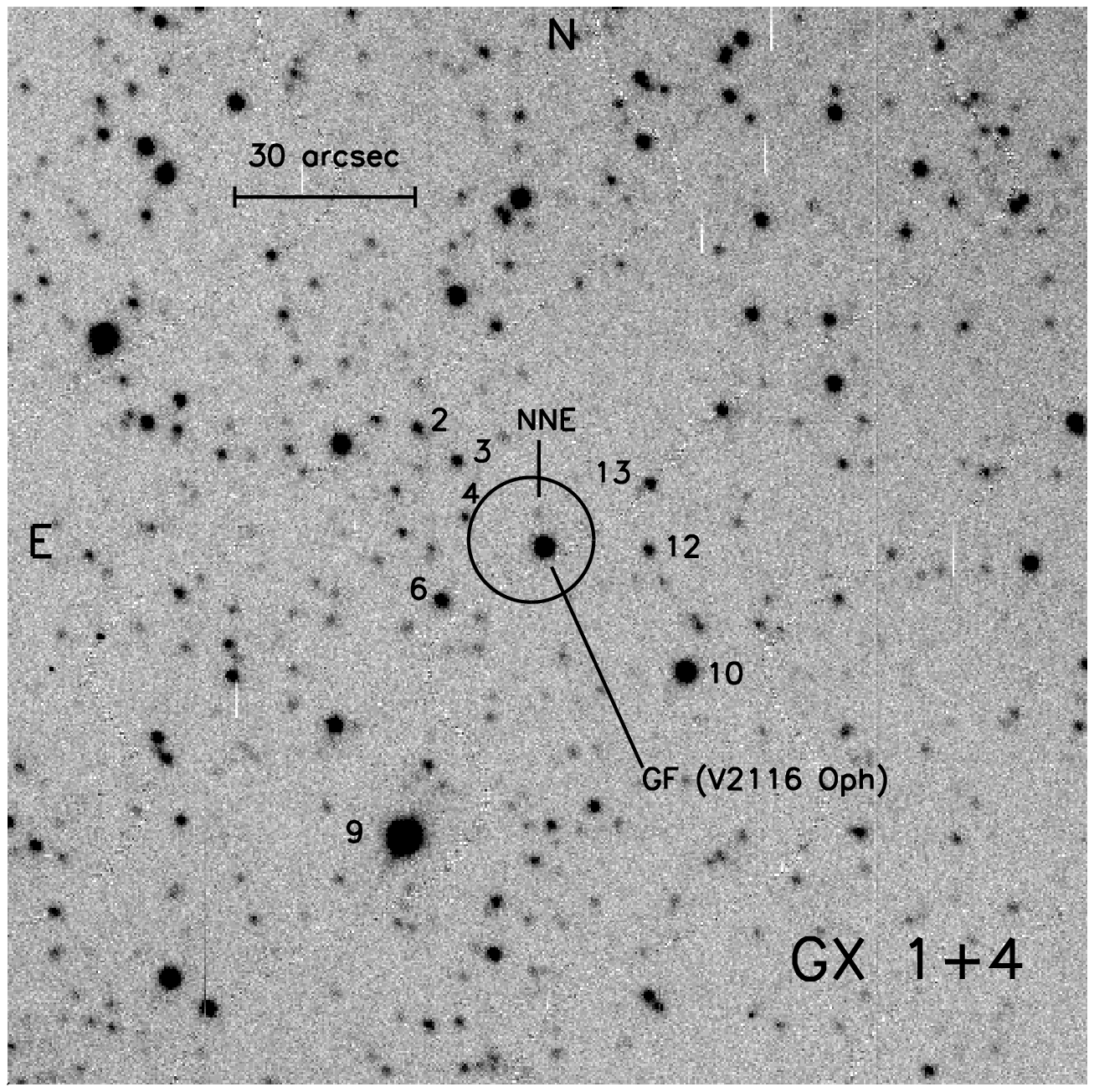}}
\end{figure}

\pagebreak
\thispagestyle{empty}
\begin{figure}
\vspace{-0.5in}
\centerline{Figure 2}
\vspace{-0.5in}
\centerline{\psfig{file=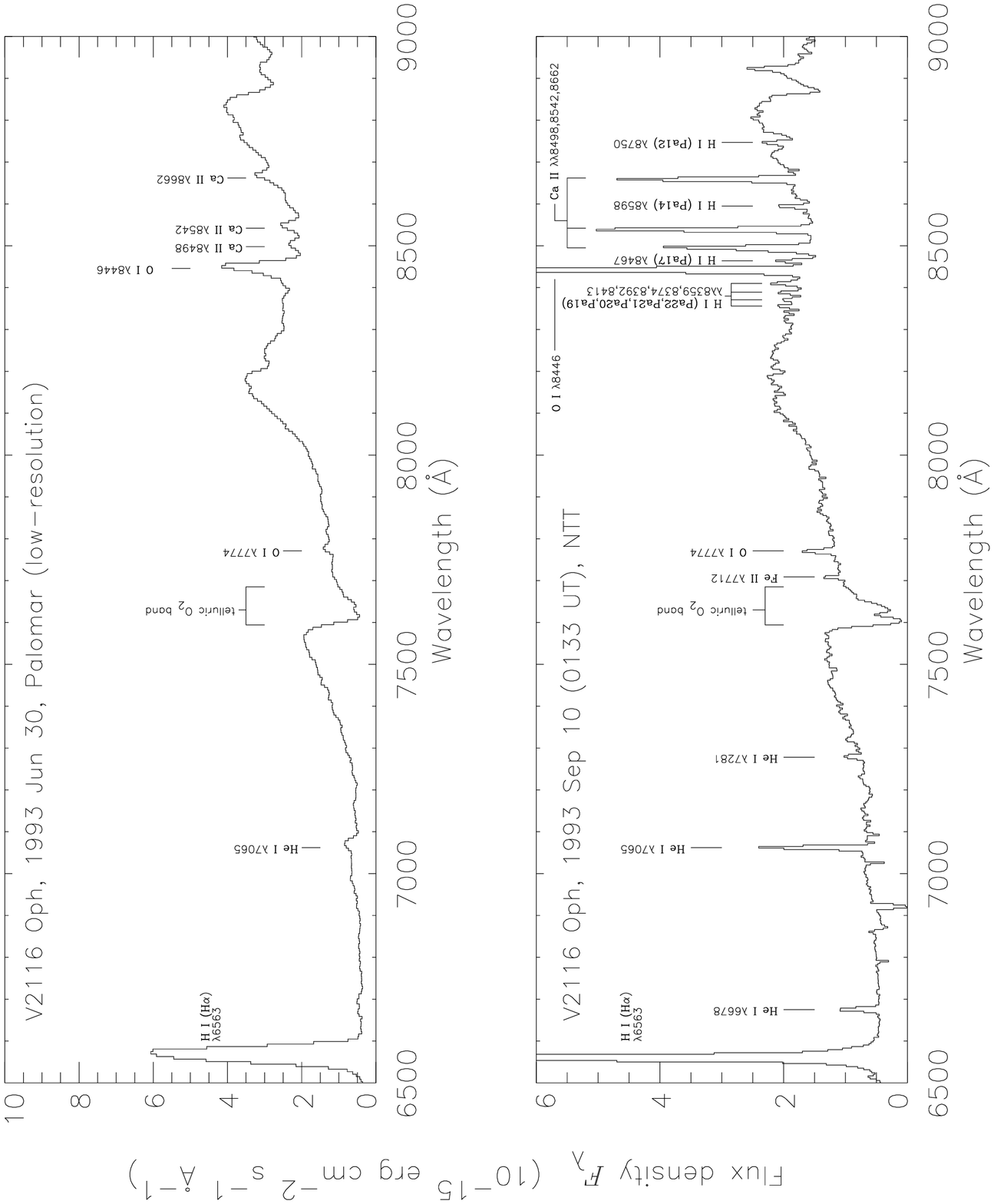,angle=180}}
\end{figure}

\pagebreak
\thispagestyle{empty}
\begin{figure}
\vspace{-0.5in}
\centerline{Figure 3}
\vspace{-0.5in}
\centerline{\psfig{file=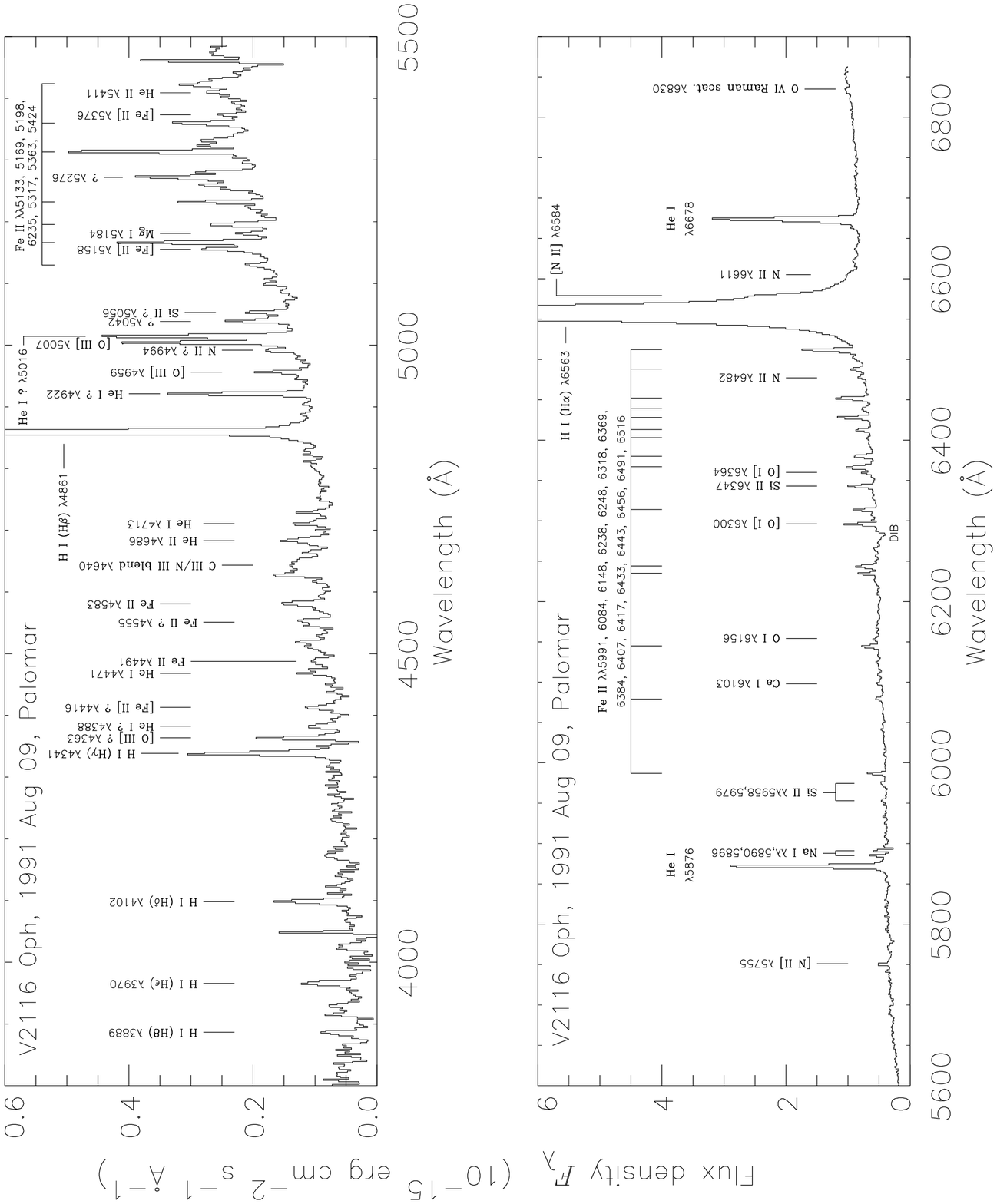,angle=180}}
\end{figure}

\pagebreak
\thispagestyle{empty}
\begin{figure}
\vspace{-0.5in}
\centerline{Figure 4}
\centerline{\psfig{file=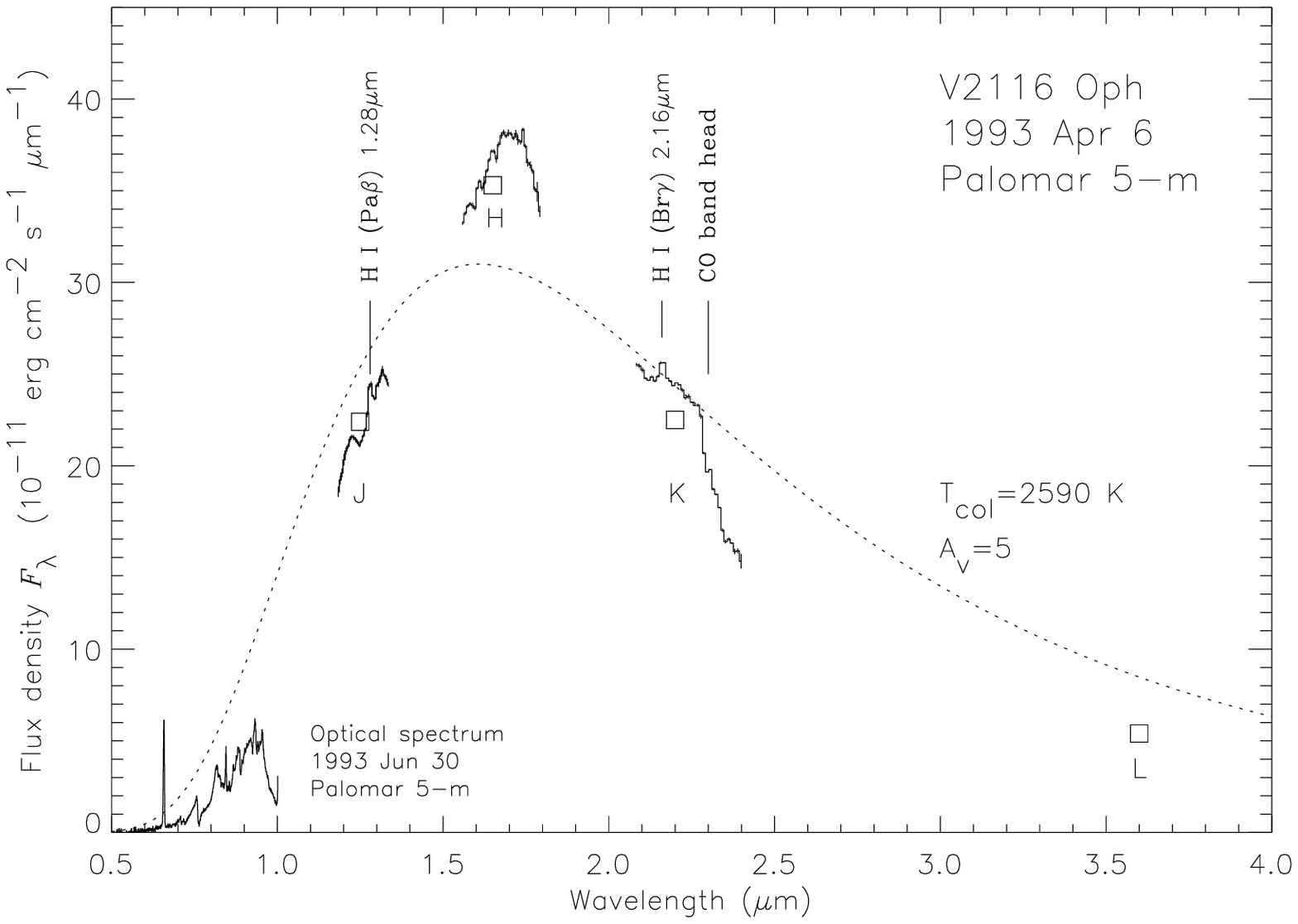}}
\end{figure}

\pagebreak
\thispagestyle{empty}
\begin{figure}
\vspace{-0.5in}
\centerline{Figure 5}
\vspace{-0.5in}
\centerline{\psfig{file=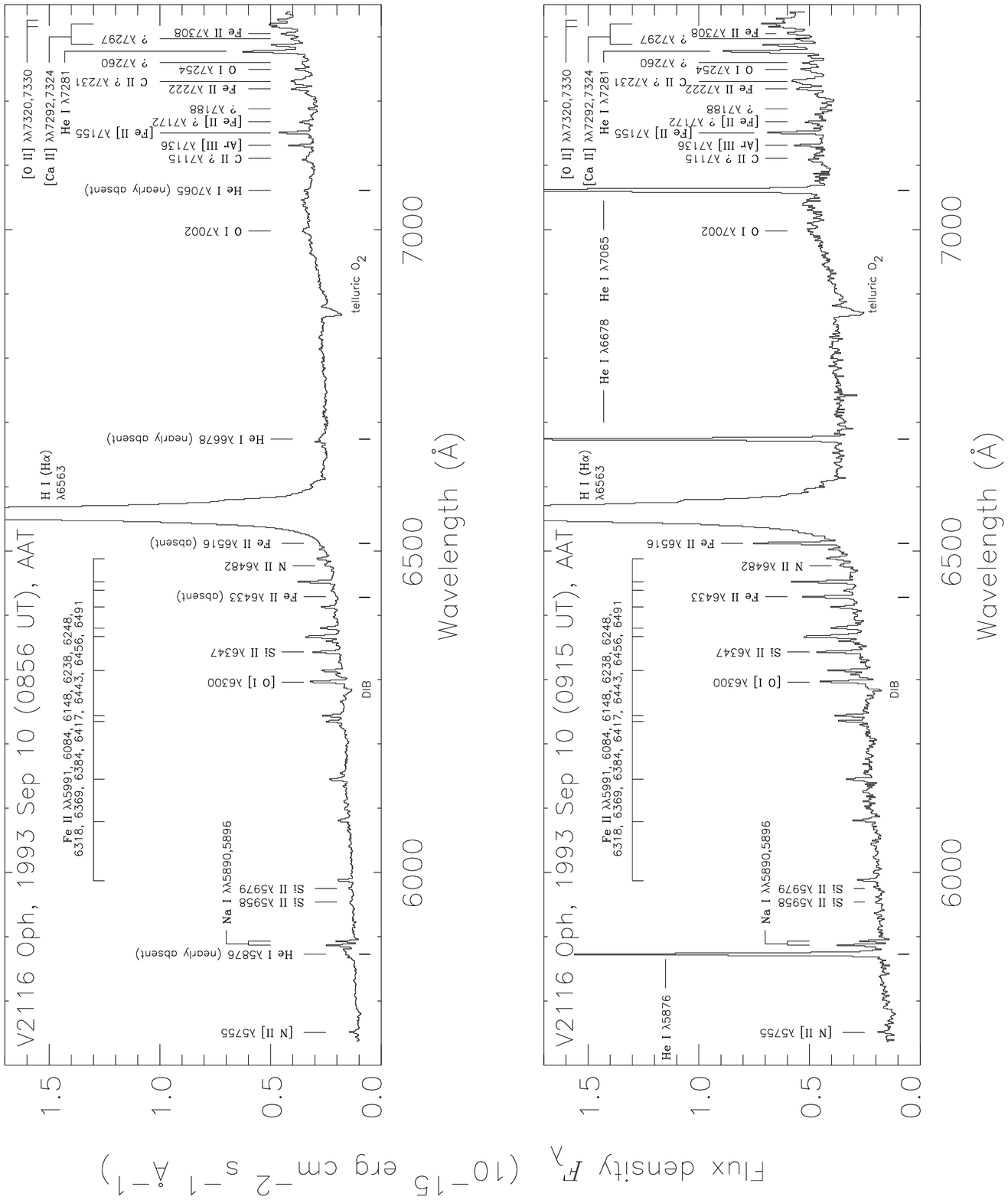,angle=180}}
\end{figure}

\pagebreak
\thispagestyle{empty}
\begin{figure}
\vspace{-0.5in}
\centerline{Figure 6}
\centerline{\psfig{file=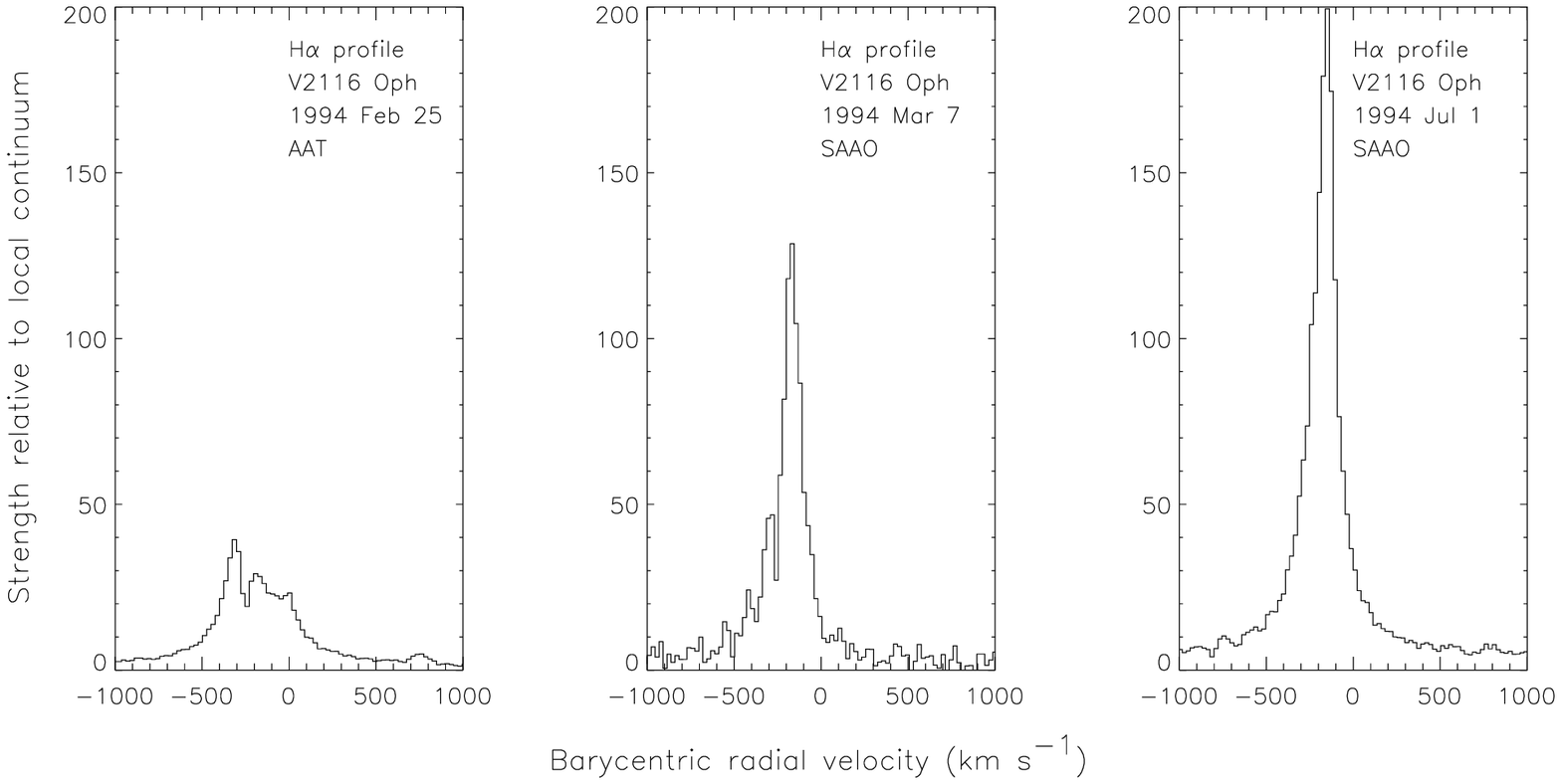}}
\end{figure}

\pagebreak
\thispagestyle{empty}
\begin{figure}
\vspace{-0.5in}
\centerline{Figure 7}
\centerline{\psfig{file=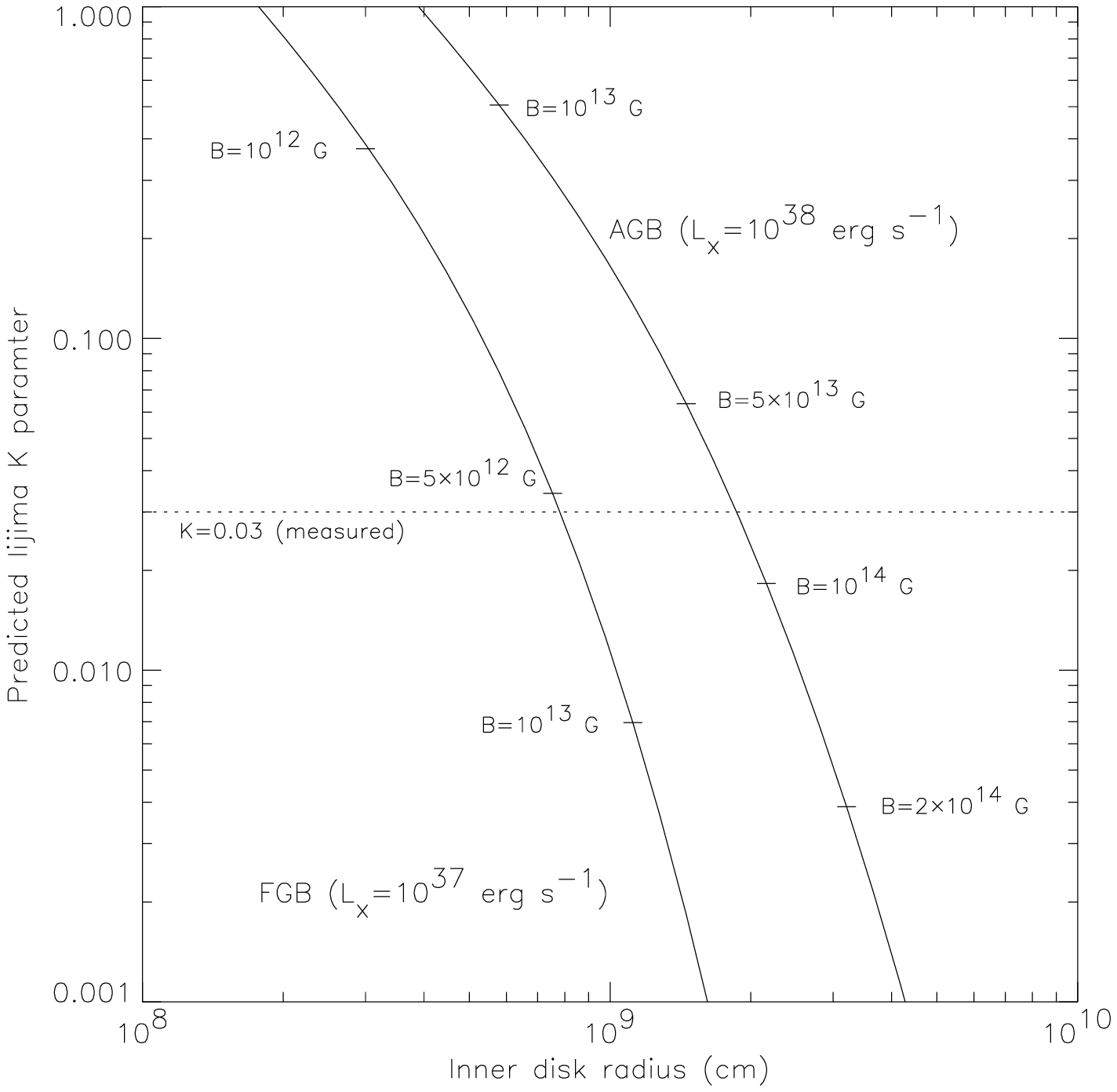}}
\end{figure}

\end{document}